\documentclass[fleqn,10pt]{wlscirep}
\usepackage[utf8]{inputenc}
\usepackage[T1]{fontenc}
\usepackage{bm}
\usepackage{xr}
\usepackage{cleveref}
\usepackage{mathtools}
\usepackage[mathscr]{euscript}
\usepackage{float}
\usepackage{url}
\usepackage{multibib}
\usepackage[ruled,vlined]{algorithm2e}
\newcites{main}{Reference}
\title{Tetris-inspired detector with neural network for radiation mapping}

\author[1,2,*,$\dagger$]{Ryotaro Okabe}
\author[1,3,4,*,$\dagger$]{Shangjie Xue}
\author[3]{Jiankai Yu}
\author[1,5]{Tongtong Liu}
\author[3]{Benoit Forget}
\author[4]{Stefanie Jegelka}
\author[6]{Gordon Kohse}
\author[6, $\dagger$]{Lin-wen Hu}
\author[1,3, $\dagger$]{Mingda Li}

\affil[1]{\textit{Quantum Measurement Group, Massachusetts Institute of Technology, Cambridge, MA 02139, USA}}
\affil[2]{\textit{Department of Chemistry, Massachusetts Institute of Technology, Cambridge, MA 02139, USA}}
\affil[3]{\textit{Department of Nuclear Science and Engineering, Massachusetts Institute of Technology, Cambridge, MA 02139, USA}}
\affil[4]{\textit{Department of Electrical Engineering and Computer Science, Massachusetts Institute of Technology, Cambridge, MA 02139, USA}}
\affil[5]{\textit{Department of Physics, Massachusetts Institute of Technology, Cambridge, MA 02139, USA}}
\affil[6]{\textit{Nuclear Reactor Laboratory, Massachusetts Institute of Technology, Cambridge, MA 02139, USA}}
\affil[*]{\textit{These authors contributed equally to this work}}
\affil[$\dagger$]{{\textit{Corresponding authors: }\href{mailto:mingda@mit.edu}{\text{rokabe@mit.edu, lwhu@mit.edu, mingda@mit.edu}}}}

\begin{abstract}
In recent years, radiation mapping has attracted widespread research attention and increased public concerns on environmental monitoring. In terms of both materials and their configurations, radiation detectors have been developed to locate the directions and positions of the radiation sources. In this process, algorithm is essential in converting detector signals to radiation source information. However, due to the complex mechanisms of radiation-matter interaction and the current limitation of data collection, high-performance, low-cost radiation mapping is still challenging. Here we present a computational framework using Tetris-inspired detector pixels and machine learning for radiation mapping. Using inter-pixel padding to increase the contrast between pixels and neural network to analyze the detector readings, a detector with as few as four pixels can achieve high-resolution directional mapping. By further imposing Maximum a Posteriori (MAP) with a moving detector, further radiation position localization is achieved. Non-square, Tetris-shaped detector can further improve performance beyond the conventional grid-shaped detector. Our framework offers a new avenue for high quality radiation mapping with least number of detector pixels possible, and is anticipated to be capable to deploy for real-world radiation detection with moderate validation. 

\end{abstract}
\begin{document}
\flushbottom
\maketitle
\thispagestyle{empty}

\section*{Introduction}
Since the Fukushima nuclear accident in 2011 till the recent risk at Zaporizhzhia nuclear power plant, there is an increasing global need calling for improved radiation detection technology, aiming to achieve high-performance radiation detection mapping with minimum impact on detectors and reduced cost. Due to the simultaneous presence of multiple radiation-interaction mechanisms, radiation detection for ionizing radiation is considerably harder than visible light. The large penetration depth of radiation, such as hard X-ray, $\gamma$-ray, and neutron, reduces the angular sensitivity of detectors and limits the majority of radiation detection efforts to focus on counting or spectra acquisition rather than their directional information. The challenge on acquiring directional radiation information further triggers additional difficulties in performing source localization, that to determine the position distributions of radiation sources\cite{connor2016airborne,lazna2017multi}. In recent years, radiation localization has attracted increased interest with applications such as autonomous nuclear site inspection. Several prototypes of system configurations have been proposed including unmanned ground\cite{lazna2017multi,christie2017radiation,guzman2016rescuer}, aerial\cite{christie2017radiation,towler2012radiation,pavlovsky20183,hellfeld2019gamma} and underwater vehicles\cite{briones1994wall,mazumdar2013ball}. Despite the remarkable progress, the information extraction process of the radioactive environment are still at an early stage with further in-depth studies much needed.

In past decades, several approaches have been proposed for directional radiation detection. One approach is the High Efficiency Multimode Imager (HEMI), which can be used to detect and locate $\gamma$-ray and X-ray radioactive sources \cite{amman2009detector,caroli1987coded,galloway2011simulation,vetter2018gamma}. A typical 
HEMI consists of two layers of CdZnTe (CZT) detectors, the first layer has a randomly arranged aperture for coded aperture imaging and the second layer is the conventional co-planar detector grid. This system requires the incident beam only come from a limited solid angle range in order to make sure the beam passes through the aperture of the first layer in order to interact with the second layer. The traditional reconstruction algorithm requires that all the incident beams should come within the field of view. If the radiation incident from another direction, the accuracy will be affected (especially for near-field radiation). Besides, this system can only conditionally detect multiple sources, usually in the case that the sources come from different isotopes and can be distinguished by energy. In this scenario, the detection with multiple sources can be reduced to single source detection by only considering the count of events within an energy range. However, in real-world applications, different sources are not necessarily distinguishable in the energy spectrum. Besides HEMI, another approach for directional radiation detection is realized by using single pad detectors separated by padding material\cite{hanna2015directional}
i.e., self-shielded method. Radiation sources from different directions and distances can result in different patterns of intensity distribution over detector arrays. Because of the inaccuracy of the model caused by misalignment and manufacture error of detector and shielding material, it is challenging to extract information from detector data via a traditional method such as non-linear fitting. Also, the traditional method is often most efficient in single sources with reduced efficiency in multiple sources. As for radiation localization and mapping, inspired by the widespread interest in Simultaneous Localization and Mapping (SLAM)\cite{cadena2016past} techniques, several works using non-directional detectors\cite{pavlovsky20183,hellfeld2019gamma} or HEMI\cite{vetter2018gamma} for radiation source localization and mapping have been presented.

\begin{figure}
\includegraphics[width=\textwidth]{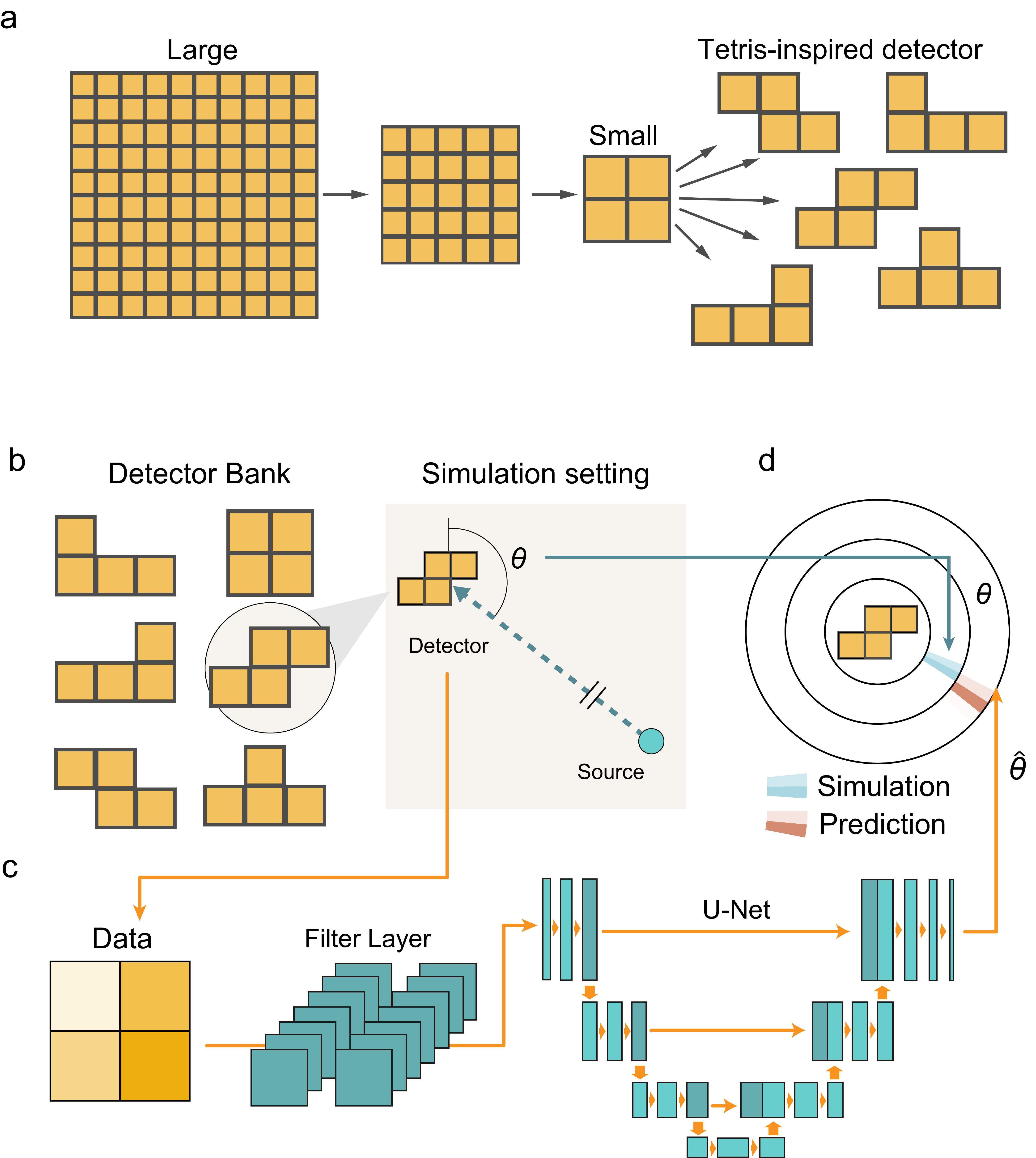}     
\caption{\textbf{Overview of Tetris-inspired radiation mapping with neural networks.}   \\
\textbf{a.} The geometrical setting of the radiation detectors. Instead of using a detector with large square grid, here we use small $2 \times 2$ square and other Tetromino shapes. Padding material is added between each pixel to increase contrast. \textbf{b-d.} The workflow for learning the radiation directional information with Tetris-shaped detector and machine learning. \textbf{b.} Monte Carlo simulation is performed to generate the detector readings for various source directions. \textbf{c.}  The detector's readouts are embedded to a matrix of filter layers for better distinguishing far-field and near-field scenarios. The embedded data then goes through a deep U-net. \textbf{d.} The predicted direction of radiation sources from the U-net (brown) with predicted angular $\hat{\theta}$ is compared to the ground-truth Monte Carlo simulations (blue) with true angular data $\theta$. The prediction loss is calculated by comparing the pairs ($\theta$, $\hat{\theta}$). 
}
\label{overview}
\end{figure}

In this work, we propose a radiation detection framework using a minimal number of detectors, combining Tetris-shaped detector with inter-pixel paddings, along with a deep-neural-network-based detector reading analysis.  Fig. \ref{overview} shows the overview of our framework. We demonstrate that detectors comprised of as few as four pixels, augmented by the inter-pixel padding material to intentionally increase contrast, could extract directional information with high accuracy. Moreover, we show that the shapes of the detectors do not have to be limited to a square grid. Inspired by the famous video game of Tetris, we demonstrate that other shapes from the Tetrominoes family, in which the geometric shapes are composed of four squares, can have potentially higher resolution (Fig. \ref{overview}a). For each shape of the detector, we generate the simulation data of the detector's input from radiation sources using Monte Carlo (MC) simulation (Fig. \ref{overview}b). Fig. \ref{overview}c shows the machine learning model we trained to predict the direction of radiation sources. Using the filter layer and the deep U-net convolutional neural networks, we establish the model to predict the radiation source direction from the detected signal. As Fig. \ref{overview}d illustrates, we compare the ground-truth label of the radiation source direction (blue) with the predicted direction (brown). By using Wasserstein distance as the loss function (see \hyperref[Methods]{Methods} for details), the model can achieve high accuracy of direction estimation. As an application of the directional detector, additional Maximum a Posteriori (MAP) has been implemented to a moving detector so that we can further estimate the spatial position of the radiation sources. Throughout this work, we limit the discussion to 2D since it is sufficient in many realistic scenario and leave the 3D discussion for future studies. 

\section*{Results}

\subsection*{Directional Prediction with Static Detectors}
\begin{figure}
\includegraphics[width=\textwidth]{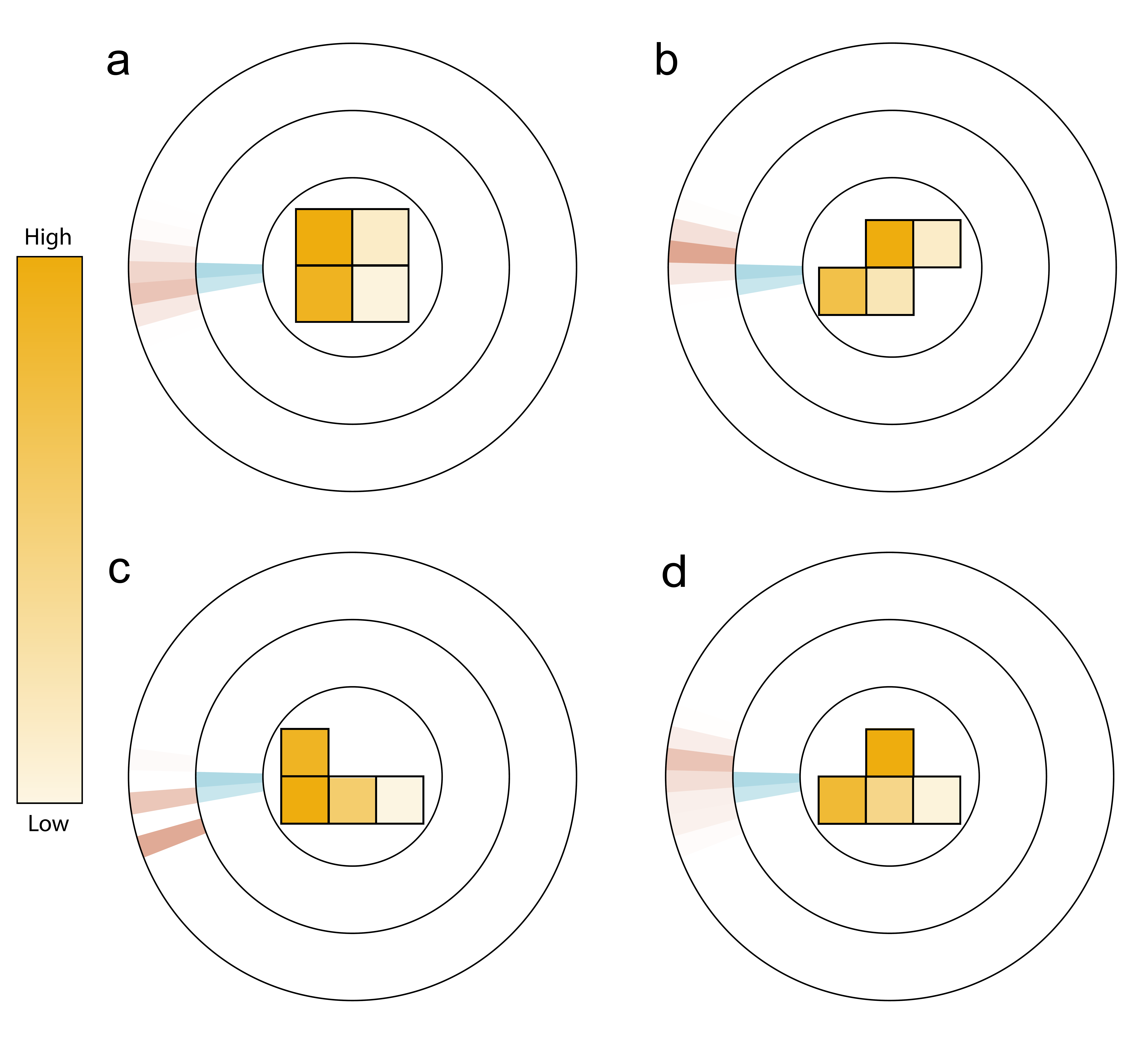} \\
\caption{\textbf{Directional radiation detection with prediction with static Tetris detectors.} In all figures, the shape in the center represent the corresponding Tetris shape, and the yellow color in each of Tetris-shaped detector indicates the detector reading, where more intense colors indicate higher signals. In all cases, the radiation is a beam of $\gamma$-ray coming from a sourcee with $d=583 cm$, $\theta=250^{\circ}$. The finite prediction of directions by the ground-truth MC simulation and the prediction are represented as blue and brown, respectively. Overall, the S-shaped Tetris detector shows consistently best performance compared to other shapes. } 
\label{result_static}
\end{figure}

First, we train the machine learning model so that the static detectors can detect the direction from which the radiation comes from. We use OpenMC\cite{romano2015openmc} package for MC simulation of the radiation detector receiving the signal from an external radiation source (more details in \hyperref[Methods]{Methods} Section). We assume that the detector pixels are composed of CZT detectors with pixel size $1cm \times 1cm$, slightly larger than the current crystals but still much smaller than the $5$ meters of source-detector distance. The inter-pixel padding material is chosen to be $1mm$-thick lead empirically, which is thick enough to create contrast but not absorb the photon in the $\gamma$-ray range. Throughout this study, we assume that the incident beam energy is $\gamma$-ray of $0.5MeV$, which is the realistic energy from pair production and comparable to many energy $\gamma$-decay energy levels. Given the energy resolution from CZT detector, radiations with other energies are also expected to be resolvable, even though here we only focus on the directional mapping where only counting matters. More details on the data preparation, normalization, neural network architectures and training procedures are shown in \hyperref[Methods]{Methods} Section. We evaluate the prediction accuracy of detectors which comprise with four detector configurations: $2 \times 2$ square grid, Tetrominos of S-, J-, T-shapes. The I-shaped Tetris detector array is not presented since it does not show performance good enough for directional mapping. The main results of the predicted radiation direction for the four Tetris-inspired detectors are illustrated in Fig. \ref{result_static} and summarized in Table \ref{tetris}. While the S-shape detector worked with the smallest prediction followed by $2 \times 2$ square, J- and T-shapes, all of the four types of detector could work enough to know the direction of the radiation source with about $1$-deg($^{\circ}$) accuracy. 

Figure \ref{result_static} shows detector readouts  and typical angular distributions predicted by neural networks in polar plots. The blue and brown color represent the ground-truth from MC and the neural network prediction, respectively. Figure \ref{result_static}a is the case with $2 \times 2$ square-grid detector , showing a typical predictive power of the radiation. The performance of other Tetris-inspired detector shapes, including S-, L- and T-shapes, are shown in Fig. \ref{result_static}b-d. By comparing with different Tetris shapes, we can see that there is a generic trend that S-shaped Tetris consistently shows best performance, while the T-shaped Tetris is the least accurate. This can be intuitively understood from a symmetry analysis. For instance, for incident radiation from the "north" with $\theta=90^{\circ}$, the left and right pixels and padding materials in the T-shape receives identical signals from radiation sources, which reduces the number of effective pixels and padding materials. Although the square Tetris in \ref{result_static}a has higher symmetry than others, it also contains four pieces of inter-pixel padding materials, in contrast to other cases with three pieces. Such analysis may also apply to the I-shaped detector array given its high symmetry.

\begin{table}[ht!]
\label{tetris}
\begin{center}
\caption{Ablation on neural network backbones with different scenarios $2 \times 2$ square grid and 3 types of Tetrominos), evaluated by the Wasserstein distance of angular distributions.}\label{tetris}
\begin{tabular}{l*{5}{c}r}
            & 2x2 & \textbf{S-shape} & J-shape & T-shape \\
\hline
Error (deg) & 0.959 & \textbf{0.864} & 1.859 & 2.001   \\
\hline
\end{tabular}
\end{center}
\end{table}

\subsection*{Positional Prediction with Moving Detectors and Maximum A Posteriori (MAP) Estimation}

\begin{figure}
    \centering
    \includegraphics[width=0.8\textwidth]{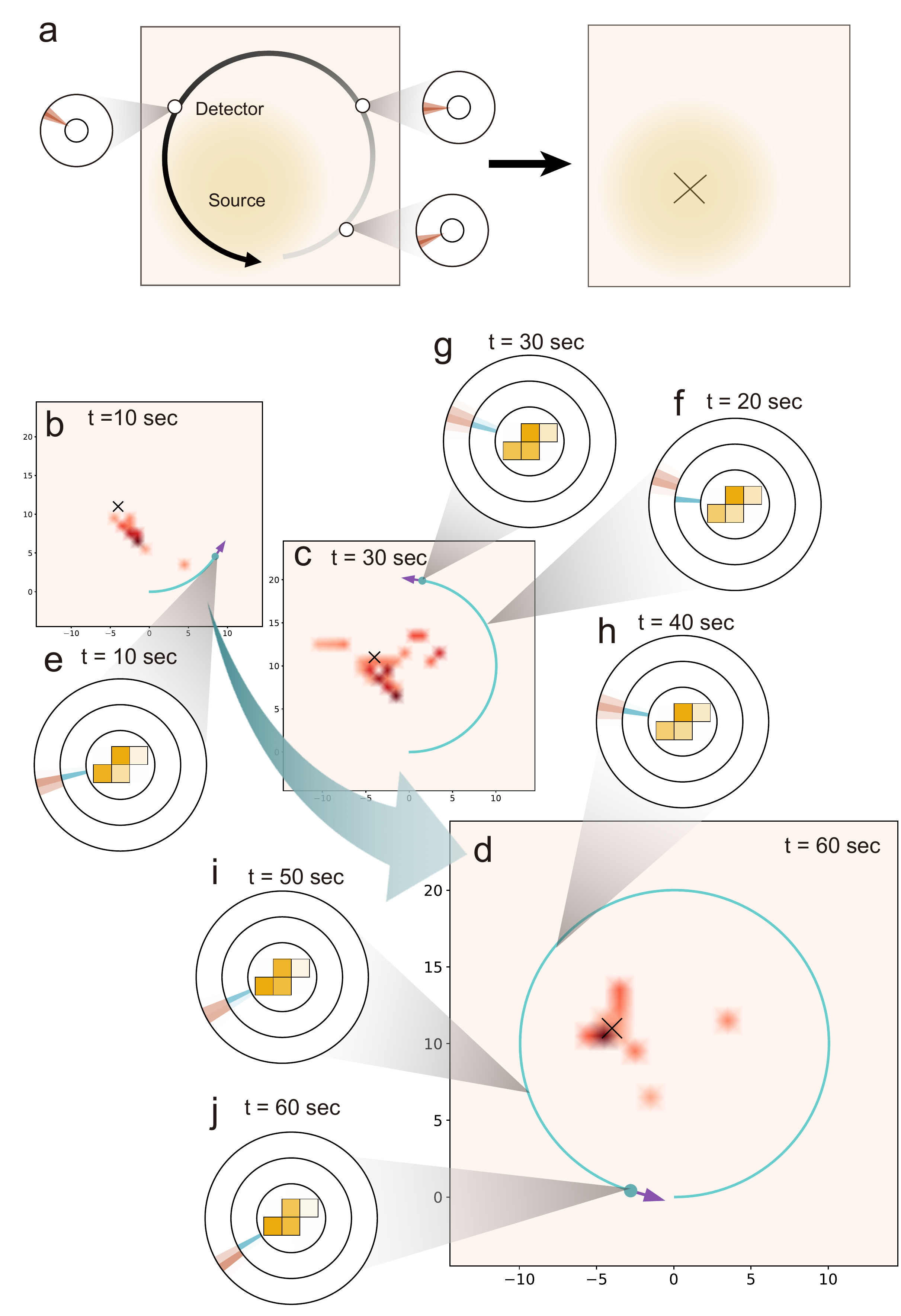}
    \caption{\textbf{Radiation mapping and position determination with a T-shape Tetris-inspired detector.} \\
    \textbf{a.} By acquiring detector readings at each spatial position during the circular motion, the position of the radiation can be gradually optimized through MAP. \textbf{b-d.} The process to map the radiation source at a few representative time at $t=10, 30, 60 \mathrm{s}$, respectively. The "$\times$" symbol on the maps shows the ground-truth location of the radiation source. The areas colored with intense red color indicate a high probability where the radiation source is located. The purple arrows indicate the front side of the radiation detector. \textbf{e-j.} The detector's input signals and the predicted directions of the radiation sources at $t=10, 20, 30, 40, 50, 60 \mathrm{s}$. Both the input signal and the diagram of the angles are visualized in the detector frame. The top side represents the front side of the moving detector. Check the radiation mapping process in Supplementary Movie 1.}
    \label{mapping}
\end{figure}

In real-world applications of radiation mapping, it would be highly desirable to go beyond the directional information but also the position of radiations. Here, a method based on Maximum A Posteriori (MAP) estimation is proposed in order to generate the guessed distribution of radiations through the motion of detectors. The workflow is as follows: first, the detector readout is simulated by MC given the detector's initial position and orientation, just as the case for the static detector. Second, the detector begins to move in a circular motion. The schematics are shown in Fig. \ref{mapping}. It is worthwhile mentioning that the particular detector face that aligns with the detector moving direction does not matter much, since the detector facing any direction is already a valid directional detector that is sensitive to radiations coming from all directions (Supplementary Information V). In other words, even if the detector is rotated intentionally or accidentally during the circular motion, the final results are still robust. Third, at each instantaneous time step during the detector motion, the predicted source direction is calculated based on the deep U-net model, just like the static detector case. Finally, the radiation source location is estimated via MAP based on the series of neural-network-inferred detector direction data at different detector positions. In an ideal case for one single isotropic radiation source, as few as two detector spatial positions are just enough to locate the source position (as the intersection of the two rays along the directions in the two detector positions), and the circular motion and MAP are implemented for more complicated radiation profile mapping. To improve the performance, we set a threshold for visualizing the radiation map. Our radiation maps are normalized by the highest probability of the presence of the radiation sources in the area of interest. We set the threshold as $0.3$ and made every value on the area of interest that is lower than this threshold as zero. This procedure enables us to visualize the mapping result clearer.

 Figs. \ref{mapping}b-j shows the dynamical process of radiation mapping and position determination. The detector geometry is the same as before, and the radius of motion is chosen randomly distributed from 0.5m to 5m. Figs. \ref{mapping}b-d show the inferred radiation mappings at three different time steps $t=10 \mathrm{s}, 30 \mathrm{s}$ and $60 \mathrm{s}$ at the beginning, half-circle, and close to the end of the circular motion. 
The ground-truth location of the radiation source is shown as the black cross in all three figures. At the early $t=10 \mathrm{s}$, there is not sufficient information for MAP to estimate the radiation position, and the estimation (red lines in Fig. \ref{mapping}b) has a ray shape that acts more like directional mapping. After $30 \mathrm{s}$, the MAP estimation is improved, though the estimated radiation is located at a broader spatial area rather than the ground truth. Finally, the detector could complete the mapping process with sufficient accuracy to point out the position of the radiation source (Fig. \ref{mapping}d). The detector's readouts and the predicted directions at each time step $(\mathrm{t}=10, 20, 30, 40, 50, 60 \mathrm{s})$ are illustrated as Figs. \ref{mapping}e-j. Radiation mapping results with other Tetris-inspired detectors are shown in Supplementary Information with additional Supplementary Videos showing the moving detector scenarios.  

\begin{figure}
    \centering
    \includegraphics[width=0.9\textwidth]{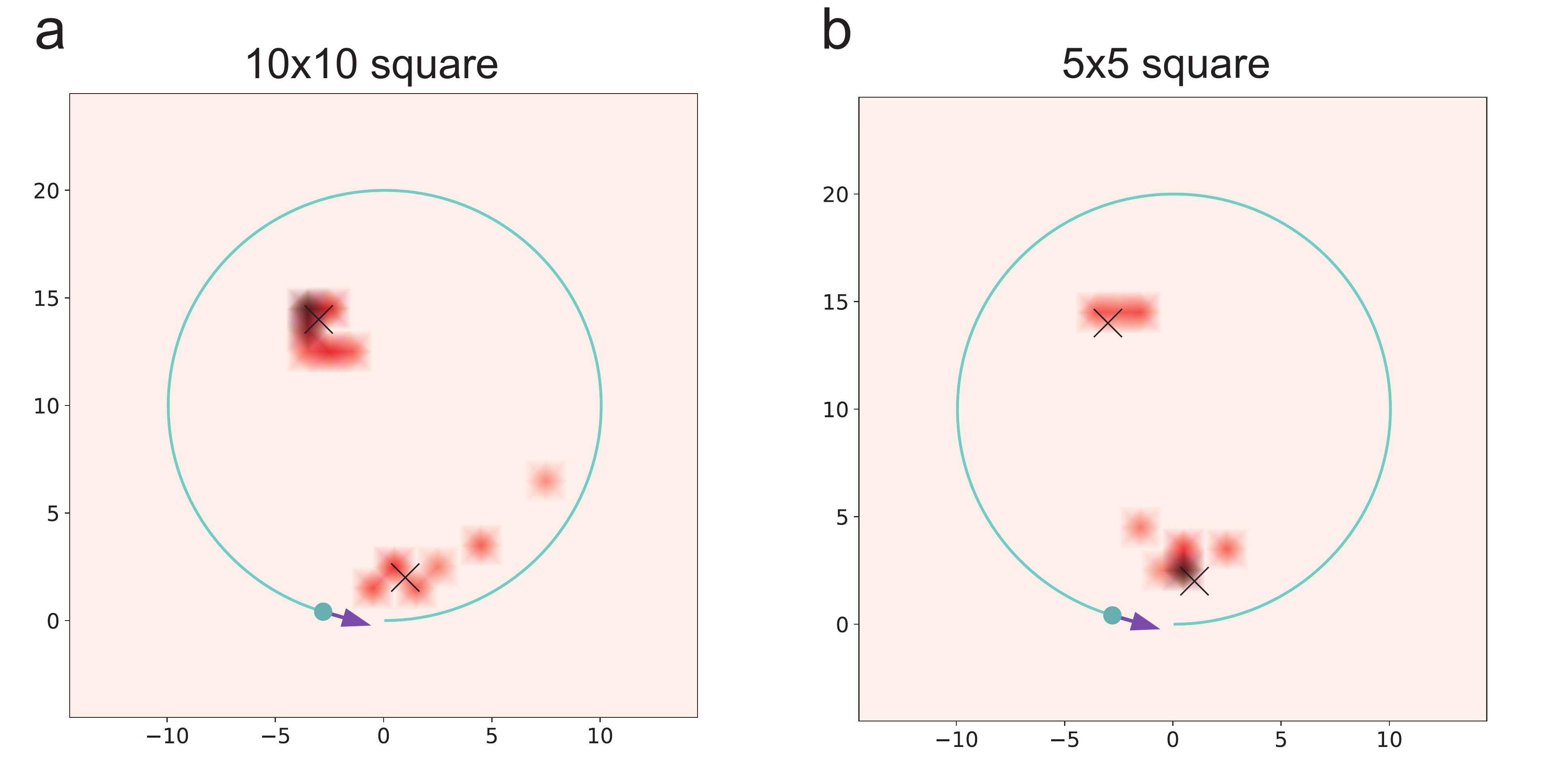}
    \caption{\textbf{Radiation mapping of two radiation sources with a moving detector.}    \\
Two radiation sources are placed in the space (shown as the two black crosses of "$\times$"). The detector is moving in a circular motion around the sources (blue circles). We use the detector of $10 \times 10$ grid (in \textbf{a.} ) and $5 \times 5$ grid (in \textbf{b.}).  
}
    \label{2sources_thhreshold}
\end{figure}

When performing a realistic radiation mapping, the area of interest may contain multiple radiation sources, which increases the level of difficulties of radiation mapping. To tackle this challenge, we further study the radiation distribution map includes multiple radiation sources (Fig. \ref{2sources_thhreshold} and Supplementary Information). We can see good agreement can be achieved for two radiation sources. However, we would like to point out that more detector pixels such as $10 \times 10$ (Fig. \ref{2sources_thhreshold}a) or $5 \times 5$ (Fig. \ref{2sources_thhreshold}b) grids are used, since the $2 \times 2$ square-grid detector does not show adequate performance even after extensive training.

\section*{Discussion}
The conventional detector configuration has a grid structure vertically facing the source of detection, where each detector pixel receives the radiation signal with a slightly different solid angle. In this work, we propose an alternative detector configuration with a few features. First, the detector grid is placed horizontally within the plane instead of vertically facing the source. Second, additional thin padding layers are padded between detector pixels, i.e., the contrast between pixels are not only created by incident angles but also enhanced by padding layers that are good absorption layers of radiation. Third, machine learning algorithms are implemented to analyze the detector reading, demonstrating great promise to reduce the need of detector pixel numbers and thereby reduce the cost of fabrication and deployment. Fourth, non-conventional Tetris-shaped detector geometry is proposed beyond the square grid, which can lead to more efficient use of pixels with improved resolution, particularly for the S-shaped Tetris detector. Finally, we demonstrate the possibility to locate the positions of radiation sources in a moving detector scheme through MAP. 

Despite these initial successes, we believe the configuration proposed in this work is still at its infancy. A number of refined works are foreseeable. As a starter, the entire work is based on MC simulations. Even if we aim to simulate a realistic environment as much as possible, it would be highly desirable to have experimental validations. In addition, although the 2D configuration can represent a number of realistic scenarios, such as radiation sources are far away from the detector but still close to ground level, it would still be an interesting problem to study 3D configuration, possibly with 3D detectors like Rubic-shaped detector cubes. Finally, a number of improvements, such as moving radiation sources and energy spectra of radiation, may be feasible with more advanced approach like reinforcement learning. Our work represents one step that leverages the detector pixels and shapes with machine learning toward radiation detection with reduced complexity and cost. \\

\section*{Code Availability Statement} 
 The source code is available at (\url{https://github.com/RyotaroOKabe/radiation_mapping.git}). 

\section*{Data Availability Statement} 
The supplementary movies showing the radiation mapping processes are available. \\

\section*{Acknowledgements}
RO, TL, GK, and ML acknowledge the support from U.S. Department of Energy (DOE), Advanced Research Projects Agency-Energy (ARPA-E) DE-AR0001298. RO and ML are partly supported by DOE Basic Energy Sciences (BES), Award No. DE-SC0021940. ML acknowledges the Norman C Rasmussen Career Development Chair, the MIT Class of 1947 Career Development Chair, and the support from Dr. R. Wachnik. \\

\section*{Methods}
\label{Methods}
\subsection*{Monte Carlo Simulation of Static Detector and Data representation}\label{sec:openmc_-simulat.ion}
The training data, in other words, the intensity measured by each detectors is simulated by MC Simulation based on the principle of radiation-matter interaction. We used an existing MC simulation package called OpenMC\cite{romano2015openmc}. Some representative results of the detector arrays is shown in Fig.\ref{result_static}. For the sake of simplicity, we temporarily assume the radiation source and the detector arrays are in the same plane. In the MC simulation, first we define the geometry of the detector. Schematic figures of detectors arrays are shown in Fig. \ \ref{overview}. The adjacent detectors (yellow) are separated by attenuation materials (black), which forms the detectors' configuration like lattices. We set the distance $d(cm)$ between the center of the detector and the radiation source. The direction of the radiation source is defined as an angle $\theta$, which is defined in clockwise direction from the front side of the detector. When we generate training data, we selected $d$ and $\theta$ at random ($d \in [20, 500], \theta \in [0, 2\pi)$) so that the neural network could get feature from radiation sources of various distances and directions. The distribution of the radiation source positions is shown in Fig. S1. After MC simulation was completed, the detector's readouts are represented as the matrix of $(\mathrm{h} \times \mathrm{w})$. For the square detector comprised with 4 detector panels, the data of $2 \times 2$ matrix was normalized so that the mean and the standard deviation are 0 and 1 respectively. For the detectors of Tetromino-shapes, the detectors' readouts are represented as $2 \times 3$ matrices. Since two sites of the matrices' 6 elements are vacant, we filled them with zero and did normalization in the same way as the square detectors. We followed the same MC simulation method as above to generate the 64 filter layers, which we explain more in detail in this section. All other parameters used in MC simulations are shown in Table. S1.

The data set $D$ is in the form of $\{\textbf{x}^{(i)}, \textbf{y}^{(i)}\}_{i\in [1, N_1]}$, where $N_1$ is the size of the data set. $\textbf{x}^{(i)}\in R^{h \times w}$ is the normalized readouts of the detector arrays h, w denotes the number of rows and columns of the detector arrays,respectively. For example,  $\mathrm{h}=\mathrm{w}=2$ for $2 \times 2$ square detector, $\mathrm{h}=2$, w $=3$ for Tetromino-shape detector. $\textbf{y}^{(i)}\in \mathbb{R}^{N_a}$ is the angular distribution of the incident radiation, $N_{a}$ is the number of sectors that are used to separate $[0,2 \pi)$. Each element in $\mathbf{y}^{(i)}$ represents the ratio of incident radiation intensity received from the direction of this sector to the total incident radiation intensity. For the point sources, the angular distribution of the radiation source is represented by the following method. For an angular distribution $\mathbf{y}$ contributed by multiple point sources, let $\mathbf{y}_{j}$ represent the angular distribution contributed by the $j^{\text {th}}$ point source. The $k^{th}$ element of $\mathbf{y}_{j}$ is defined by

\begin{equation}
  y_{jk} =
    \begin{cases}
      0, & \text{if } |\theta_j - \frac{2\pi}{N}(k-1)| \geq \frac{2\pi}{N}\\
      \frac{|\theta_j - \frac{2\pi}{N}(k-1)|}{\frac{2\pi}{N}}, & \text{else}
    \end{cases}       
\end{equation}

We also have:

\[\textbf{y} = \sum_{j}\frac{I_j}{I_0}\textbf{y}_j\]

where $\theta_{j}$ denotes the incident angle of the $j^{th}$ radiation source, $I^{j}$ denotes the total incident intensity revived from the $j^{th}$ radiation source, $\mathrm{I}_ 0=\Sigma_{j} I_{j}$ s the total incident intensity revived from all radiation sources. By using this representation, we are able to accurately represent the incident direction of a point source with an arbitrary angle, with a discretized angle interval. In the experiments, $[0,2 \pi)$ is separated in to $N_{a}=64$ sectors. Fig. 2 illustrates this representation by a pie chart.

\subsection*{Deep Neural Network Architecture}
In order to extract the global patterns of the input data, a set of global filters is designed. We obtain several filters based on very high-quality simulations in some representative cases, including far-field incident (the radiation source is located at a far distance compared to the size of the detector arrays) and near-filed (the radiation source is located at a close distance) incident at different directions. As an example, near-field filters of the S-shape detector are shown in Fig. S2. Each filter has the size of $(\mathrm{h} \times \mathrm{w})$, which is set same as that of the training data. The weight of each unit is given by the readout of each single pad detector in the simulation. The output of this layer is given by:

\begin{equation}
Z_{mk} = \sum_{i}\sum_{j}x_{ij}w_mF_{mkij} + b_{mk}\
\end{equation}

where $Z_{m k}$ is the $k^{t h}$ element of the output array at channel $m \in \{1,2\}$. Channel 1 and 2 correspond to the far-field and near-field filters respectively. $x_{ij}$ denotes the input array at pixel $\mathrm{i}, \mathrm{j}, F_{m k i j}$ denotes the $(\mathrm{i}, \mathrm{j}$) element of the global filter obtained in the case that incident radiation from the $k^{th}$ sector, in far-field $(\mathrm{m}=1)$ or near-field $(\mathrm{m}=2)$ scenario. $w_{m}$ denotes a channel-wise normalization weight, and $b_{m k}$ denotes the bias for this global filter. During training, the weights of the global filter are initialized with the far field and near-field filters that we obtained from the high-quality simulations. The weights are slightly fine-tuned with a learning rate smaller than the learning rate of the other layer of the network, while the bias for each filter is trained with the learning rate same to the other layers. The filter layer is followed by an Exponential Linear Unit (ELU) activation function\cite{clevert2015fast}. The output of the global filtering layer embeds the directional information corresponding to the direction of the filter channel. It is then fed into the U-shape network to extract the directional information.

In the neural network, the input data is normalized $(\mathrm{h} \times \mathrm{w})$ detectors readout. However, it is essentially different from images captured by cameras. For the processing of image which is measured from visible light, convolution neural networks (CNNs) are widely used\cite{lecun2015deep}. A convolution layer is used for extracting features that are presented as localized patterns. However, due to the penetration properties of several kinds of radiation, such features are presented as global patterns, which is different from the imaging of visible light. Therefore a novel architecture is designed for this purpose, the input data is followed by a global filtering layer with the shape $(2, 64, \mathrm{h}, \mathrm{w})$, in order to extract the global pattern. The output of this layer conveys the directional information with a size of $(2, 64)$, which corresponds to the final output, i.e. the estimated angular distribution, with a size of $(1, 64)$. In order to perform a pixel-to-pixel prediction of the angular distribution, a U-shape fully convolutional architecture which is similar to U-net\cite{ronneberger2015u} is then utilized as Fig.\ref{overview}b-d shows. Noticed that the U-Net architecture is originally developed for image segmentation\cite{ronneberger2015u}. However here the output is 1D array, and the input is viewed as 1D arrays with 2 channels. Thus we accordingly set the dimension of the U-Net, as is shown in Fig. S3. Finally, the output of the final layer feeds into a softmax layer for normalization.

In order to represent the distributional similarity between the predicted and target distribution, Wasserstein distance is proposed to be used as the loss function. It is a distance function on a given metric space between two probability distributions. As this metric is an analogy of the minimum cost required to move a pile of earth into the other, it is also known as the earth mover's distance\cite{talebi2018nima,hou2016squared,genevay2018learning}. The Wasserstein loss function is given by:

\begin{equation}
W = \min_{\pi (i, j)} \sum_{i}^{n}\sum_{j}^{n} \pi(i,j)C(i,j)\
\end{equation}
$
\begin{aligned}
\text{subject to} ~~~~~~~~~ \pi(i,j) &\geq 0, ~~~&\forall i,j \in [1,N] \\
\sum_j \pi(i,j) &\leq y_i, ~~&\forall i \in [1,N] \\
\sum_i \pi(i,j) &\leq \hat{y}_j, ~~&\forall j \in [1,N] \\
\sum_i \sum_j \pi(i,j) &=1   ~~&
\end{aligned}
$
\\
\noindent where  $\pi(i,j)$ is the transport policy which represents the mass to be transferred from state $i$ to state $j$, $y_i$ is the ground truth of the normalized angular distribution, $\hat{y_i}$ is the estimated angular distribution. $C(i,j)$ is the cost matrix representing the cost of moving unit mass from state $i$ to $j$. Our work is in a cyclic case and use the following form:

\begin{equation}
C(i,j) = {min(|i-j|, |j+n-i|, |i+n-j|)}^l\
\end{equation}

An algorithm is developed to calculate the cyclic Wasserstein distance. Particularly, the cyclic case is unrolled into an ordered case. The ring is split into a line at $n$ different units and obtain $n$ different distributions. the cost matrix in the cyclic and ordered cases. The Wasserstein distance could be computed in closed form\cite{levina2001earth,hou2016squared}:

\begin{equation}
W(p, t) = (\frac{1}{n})^{\frac{1}{l}}{\lVert  CDF(p) - CDF(t)  \rVert}_l\
\end{equation}

Where $CDF(\cdot)$ calculate the cumulative distribution of its input. Following this formula, a decycling algorithm is developed to calculate the Wasserstein distance with a cyclic cost. The algorithm is shown below:

\begin{algorithm}[H]
\SetAlgoLined
\KwIn{Array $p$ and $t$ of size $N$, }
\KwOut{Wasserstein Distance: $Dist$}
 $\mathbf{d}$ $\leftarrow$ new array of $N$\\
 \For{i $\leftarrow$ 1 \KwTo n}{
  $p_{new}$ $\leftarrow$ Concat($p[i+1$ \KwTo $N]$, $p[1$ \KwTo $i]$)\\
  $t_{new}$ $\leftarrow$ Concat($t[i+1$ \KwTo $N]$, $t[1$ \KwTo $i]$)\\
  $d[i]$ $\leftarrow$ $(\frac{1}{N})^{\frac{1}{l}}||CDF(p)-CDF(t)||_l$
 }
 $Dist$ $\leftarrow$ min($\mathbf{d}$)
 \caption{Cyclic Wasserstein Distance \label{alg:1}}
\end{algorithm}

It could be numerically verified that this algorithm enables exact calculation of $1^{st} (l=1)$ Wasserstein distance for the cyclic Wasserstein distance, given arbitrary distribution. This algorithm is differentiable and enables us to optimize the objective through back-propagation. As for evaluation in the experiments, we use $1^{st}$ Wasserstein distance which can directly represent the angle difference of the estimated and real directions. In the training process, we use the above $2^{nd}(l=2)$ Wasserstein distance as loss function since it usually converges faster with gradient descent-based optimization methods, compared to the $1^{st}$ Wasserstein distance\cite{shalev2009stochastic,hou2016squared}. \\

In the proposed model, the network is trained using Adam\cite{kingma2014adam} with a learning rate of $0.001$ for all parameters in the neural network except for the weights of the global filtering layer, whose learning rate is set to $3\times10^{-5}$. The training batch for each step is randomly selected from the training set which is based on several pristine simulation result set $D^{train}$ on one radiation source. All models randomly split the data into 90\% training (2,700 data), and 10\% testing (300 data) sets. Furthermore, we trained our models with a 5-fold cross-validation scheme.

\subsection*{Radiation Source Mapping with Maximum A Posteriori (MAP) estimation}
We set up a radiation mapping problem by considering the case that there is one point source in the environment. This problem could be treated as a Simultaneous Localization and Mapping (SLAM) problem\cite{cadena2016past}. The directional radiation detector could be viewed as a kind of sensor which could only obtain directional information, and a radiation source could be viewed as a special landmark that could only be measured by this kind of detector. We show that by treating the directional radiation detector as a sensor with only directional resolution, it could be easily integrated into the MAP optimization framework, and enables source localization. Details of how to integrate the directional radiation detector into MAP framework are presented in Fig. S4.

Here, a method based on Maximum a Posteriori (MAP) Estimation is proposed in order to generate the radiation distribution map. In addition, here we assume that the carrier for the detector has already localized itself and is only required to build the radiation map. The map is discretized into a mesh with $N_{m}$ square pixels. Let $c \in \mathbb{R}^{N_{m}}$ denotes the radiation concentration at each pixel. It could be assumed that the radiation is uniformly generated from the pixel. The measurements result $Z_{t}=I_{0} y_{t}$ denotes the incident radiation intensity coming from different directions at time $t$. It could be assumed that the measurement probability  $p(z_t|c)$ is linear in its arguments, with added Gaussian noise:

\begin{equation}
z_t =  M_t c + \delta\
\end{equation}

\noindent where $\delta \sim N\left(0, \Sigma_{\delta}\right)$ describes the measurement noise, $M_{t} \in \mathbb{R}^{N \times N_{m}}$ denotes the observation matrix at time $t$. Considering the contribution of one pixel to one direction sector of the detector, only the overlapped area can contribute to the sector, as is shown in Fig. S4, and the intensity is proportional to the overlapped area. Besides, the intensity is inverse proportional to the square of the distance between the detector and the source. Therefore, the element of $M_{t}$, or in other words the intensity contribution of the $i^{th}$ pixel to the $j^{th}$ directional sector of the detector at time $t$ can be written as:

\begin{equation}
M_{tij}=  \frac{A_{tij}}{{r_{ti}}^2}\
\end{equation}

\noindent where $A_{t i j}$ denotes the area of the overlapped region of the $i^{th}$ pixel and the $j^{th}$ sector at time $t$ (blue area in Fig. S4), $r_{t i}$ denotes the distance between the detector and the center of the pixel. According to Bayes' rule, we have:

\begin{equation}
\begin{split}
p(c|z_1, z_2, ..., z_t) &= \frac{p(z_1, z_2, ..., z_t|c)p(c)}{p(z_1, z_2, ..., z_t)} \\
&\propto p(z_1, z_2, ..., z_t|c)p(c) \\
&= p(c)\prod_{i=1}^{t}p(z_i|c)
\end{split}
\end{equation}

\noindent Here we assume that measurements at different time are conditionally independent given $c$. As we are trying to find $c$ that maximizes $p(c|z_1, z_2, ..., z_t)$, the $p(z_1, z_2, ..., z_t)$ term could be ignored as it is independent of $c$. It could be assumed that the prior term $p(c) = N(0, \epsilon I)$ is a Gaussian distribution. Then following the the Maximum a Posteriori (MAP) estimation, we have:

\begin{equation}
\begin{split}
\operatorname*{argmax}_cp(c|z_1, z_2, ..., z_t) &= \operatorname*{argmax}_c\sum_{t}\ln{p(z_t|c)} + ln{p(c)}\\
&= \operatorname*{argmin}_c\sum_{t}\lVert \bm{M}_tc - z_t\lVert^2_{\bm{\Sigma_\delta}} + \frac{1}{\epsilon^2}{\lVert c \lVert}_2^2\\
\end{split}
\end{equation}

Therefore, the radiation concentration distribution could be obtained by solving the optimization problem:
\begin{equation}
 \min_{\bm{c}} ~~ \sum_t \|| \bm{M}_t \bm{c} - \bm{z}_t ||^2_{\bm{\Sigma_\delta}} + \frac{1}{\epsilon^2} ||\bm{c}||^2_2 \\
\text{subject to} ~~ c_i \geq 0, ~~\forall i \in [0,N_m]
\end{equation}

\noindent where $\frac{1}{{\epsilon^2}}\lVert c \rVert_2^2$ term could be viewed as a regularization term. If we do not have much information regarding the prior distribution, $\epsilon$ will be a large number. This term will penalize large concentration if the measuring data is inadequate to determine the concentration (i.e. the area is not fully explored and caused very small $M_{tij}$). In practice, $\epsilon$ could be tuned by utilizing differentiable convex optimization layers\cite{agrawal2019differentiable}, in which the optimization problem could be viewed as a layer within the neural network, and error back-propagation is enabled through implicit differentiation, given predicted and ground truth data. In our demonstration case, we simply set $\frac{1}{\epsilon^2} = 0.1$ such that the regularization term is relatively small.

\bibliography{bibliography}

\end{document}


\maketitle

\newcommand{\beginsupplement}{%
        \setcounter{table}{0}
        \renewcommand{\thetable}{S\arabic{table}}%
        \setcounter{figure}{0}
        \renewcommand{\thefigure}{S\arabic{figure}}%
     }

\beginsupplement

\section{Monte Carlo Simulation Setting and Data Representation}
\begin{figure}[H]
\includegraphics[width=\textwidth]{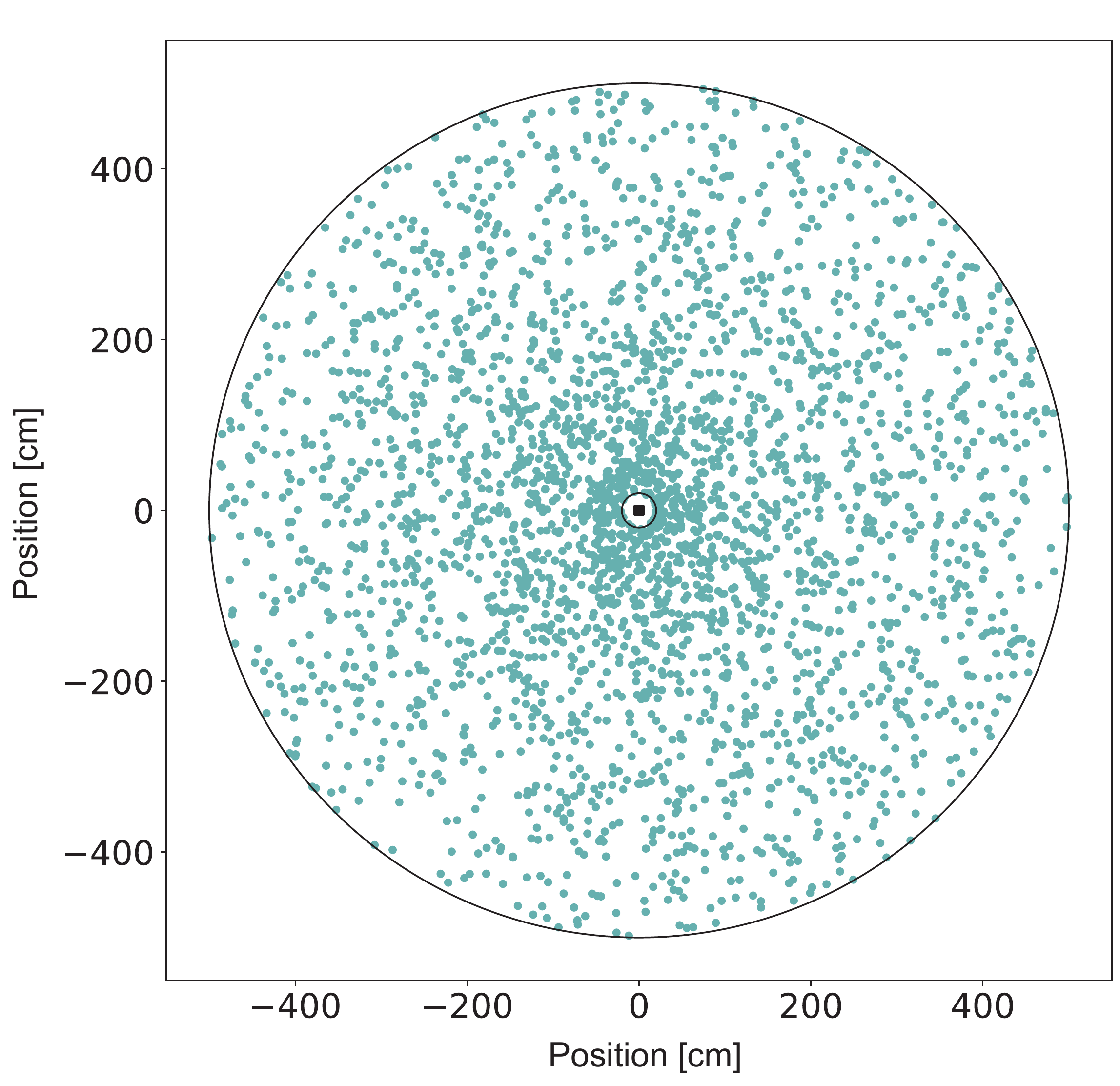}
\caption{\textbf{Visualizations of the radiation source positions of training data.} \\ 
the black square is the detector's position. Blue dots are the radiation source positions (3,000 data). The circles shows the minimum and maximum distance between the radiation sources and the detector.  
}
\label{SI_src_pos}
\end{figure}

\begin{figure}[H]
\includegraphics[width=\textwidth]{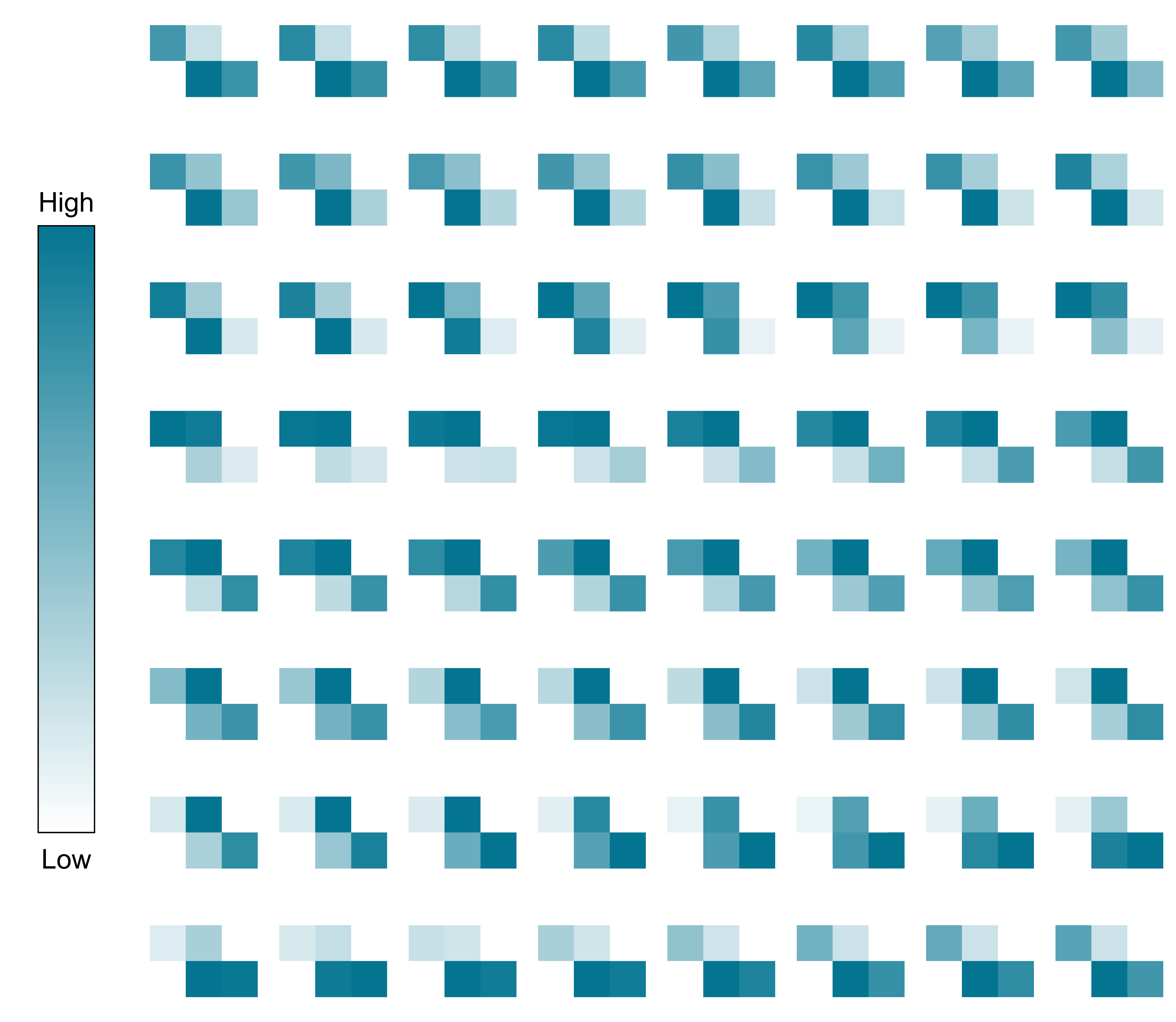}
\caption{\textbf{The near-field global filters of the S-shape configuration.}  \\ 
All of the 64 arrays represents the filters with a radiation source coming from different direction. As the color bar of the left side shows, the intense color represents higher input signals. We use the another set of 64 far-field filters to capture the direction of radiation sources. 
}
\label{SI_filterlayers}
\end{figure}

Our radiation detectors consist of two materials: CdZnTe to absorb radiation and Pd as padding layers. The density of CdZnTe and Pd are set 5.76 $g/cm^3$ and 11.35 $g/cm^3$ respectively. Table \ref{SI_params_openmc} shows the MC simulation settings to generate training data and the filter layers. Figure \ref{SI_src_pos} shows the position of the radiation sources used for training of S-shape detector. The 64 near-field filters of the S-shape detector are visualized in Fig. \ref{SI_filterlayers}.

\begin{table}[H]
\begin{center}
\caption{The parameter setting for Monte-Carlo simulations for radiation source detection.}\label{SI_params_openmc}%
\begin{tabular}{@{}llll@{}}
 & Training, Testing Data & Filter\\
\hline\hline
The number of photon particles & 50000 & 80000\\
The number of angle sectors & 64 & 64\\
radiation source energy [MeV] & 0.5 & 0.5\\
The detector-source distance [cm] & 20 $\sim$500 & near: 50, far: 500\\
The number of simulation data & 3000 & 64$\times$2 channels \\
\hline
\end{tabular}
\end{center}
\end{table}

\newpage
\section{Deep Neural Network Architecture}
\begin{figure}[H]
\includegraphics[width=\textwidth]{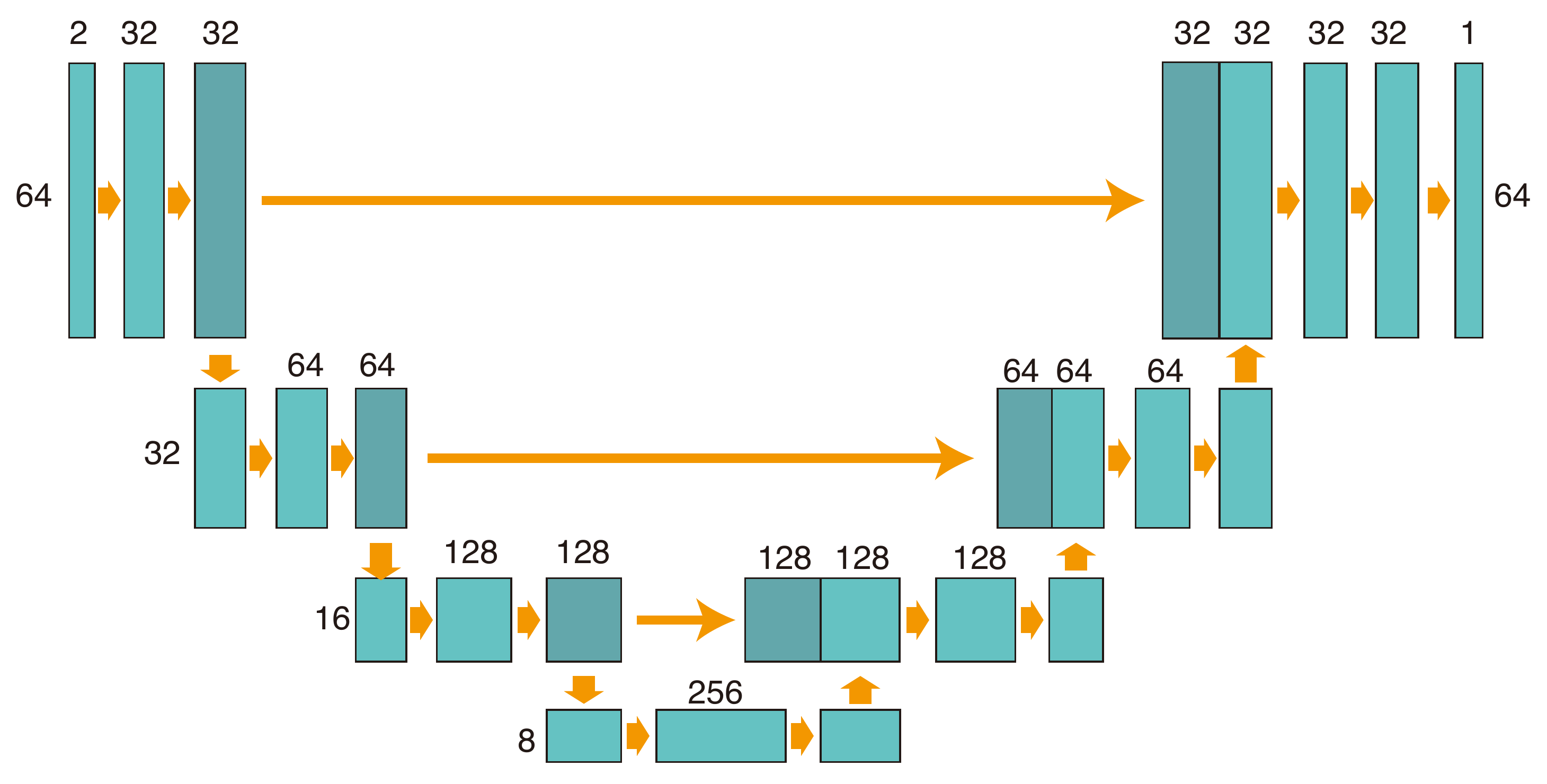}
\caption{\textbf{Model architecture of U-net convolutional neural networks.} \\ 
The number near the rectangular shapes represents the digits of matrices. The model consists of convolution, transposed convolution, max pooling and concatenation.
}
\label{SI_u_net}
\end{figure}

We implemented U-shaped neural network to predict the angular information from the detector's signal. The dimensions of each layer are shown in Fig. \ref{SI_u_net}. The input data goes through 64 filter layers of both far- and near-field channels, and embedded as $(2, 64)$ matrix. The U-net is then applied to this embedded data and output an array of 64 elements. Table \ref{SI_hparams} summarizes the hyperparameters of the U-shaped neural network. 

\begin{table}[H]
\begin{center}
\caption{The parameter setting of U-net}\label{SI_hparams}%
\begin{tabular}{ccc}
 Hyperparameter & 1-source detector & 2-source detector \\
\hline\hline
Epochs   & 200 & 5000  \\
The number of angle sectors   & 64 & 64 \\
batch size   & 256 & 256  \\
AdamW optimizer learning rate   & $1\times10^{-3}$ & $1\times10^{-3}$ \\
AdamW optimizer weight decay coefficient   & $3\times10^{-5}$ & $3\times10^{-5}$ \\
Training data size & 2700  & 2700  \\
Testing data & 300 & 300   \\
\hline
\end{tabular}
\end{center}
\end{table}

\newpage
\section{Radiation Source Mapping with Maximum A Posteriori (MAP) estimation}
\begin{figure}[H]
\includegraphics[width=\textwidth]{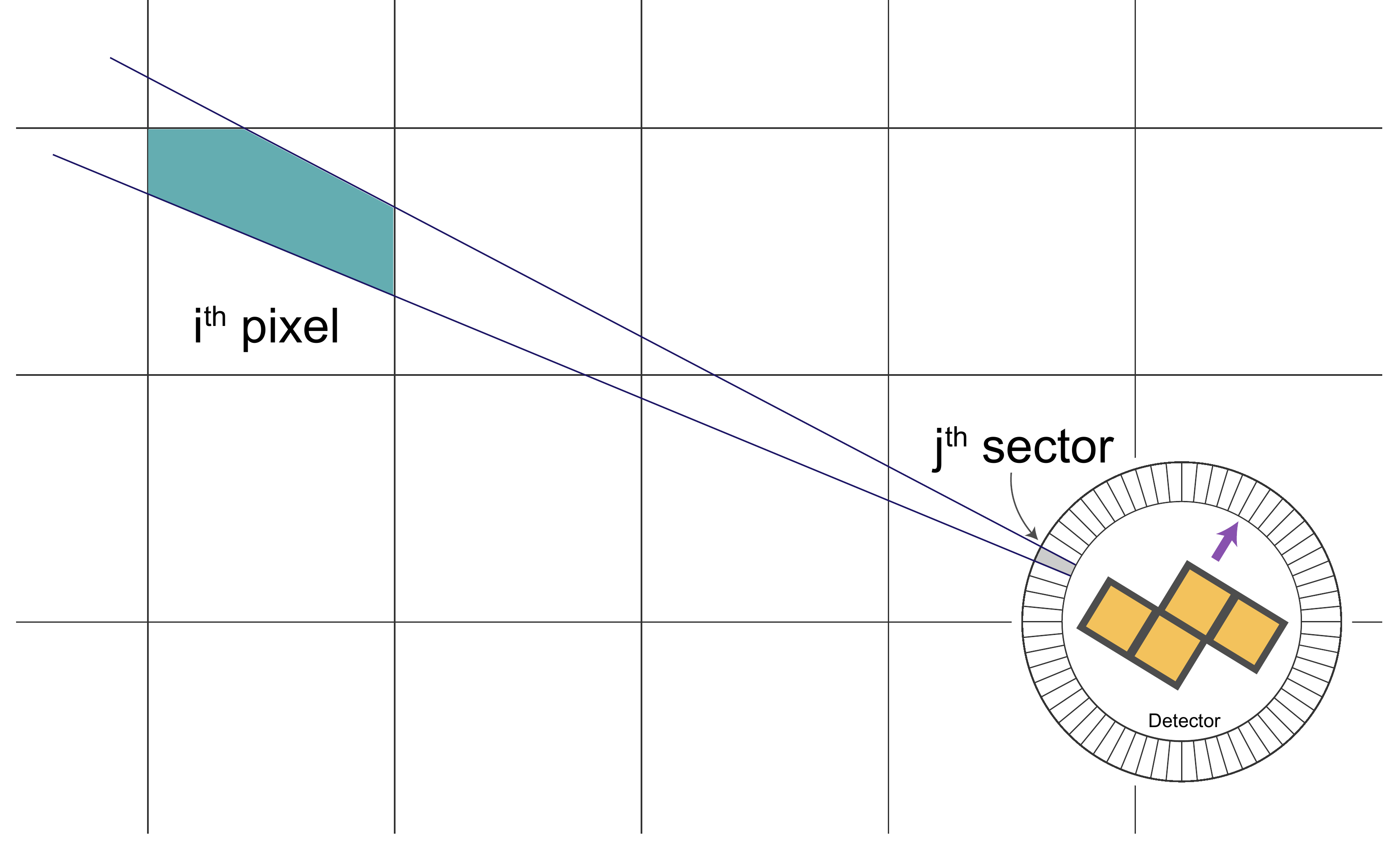}
\caption{\textbf{An illustration of the element of the observation matrix.}  \\
The purple arrow near the detector indicates the front side of the detector. The blue area on the grid indicates the overlapped region of the $i^{th}$ pixel and the $j^{th}$ sector of detection angle.
}
\label{SI_map_explain}
\end{figure}

\begin{figure}[H]
\centering
\includegraphics[height=0.85\textheight]{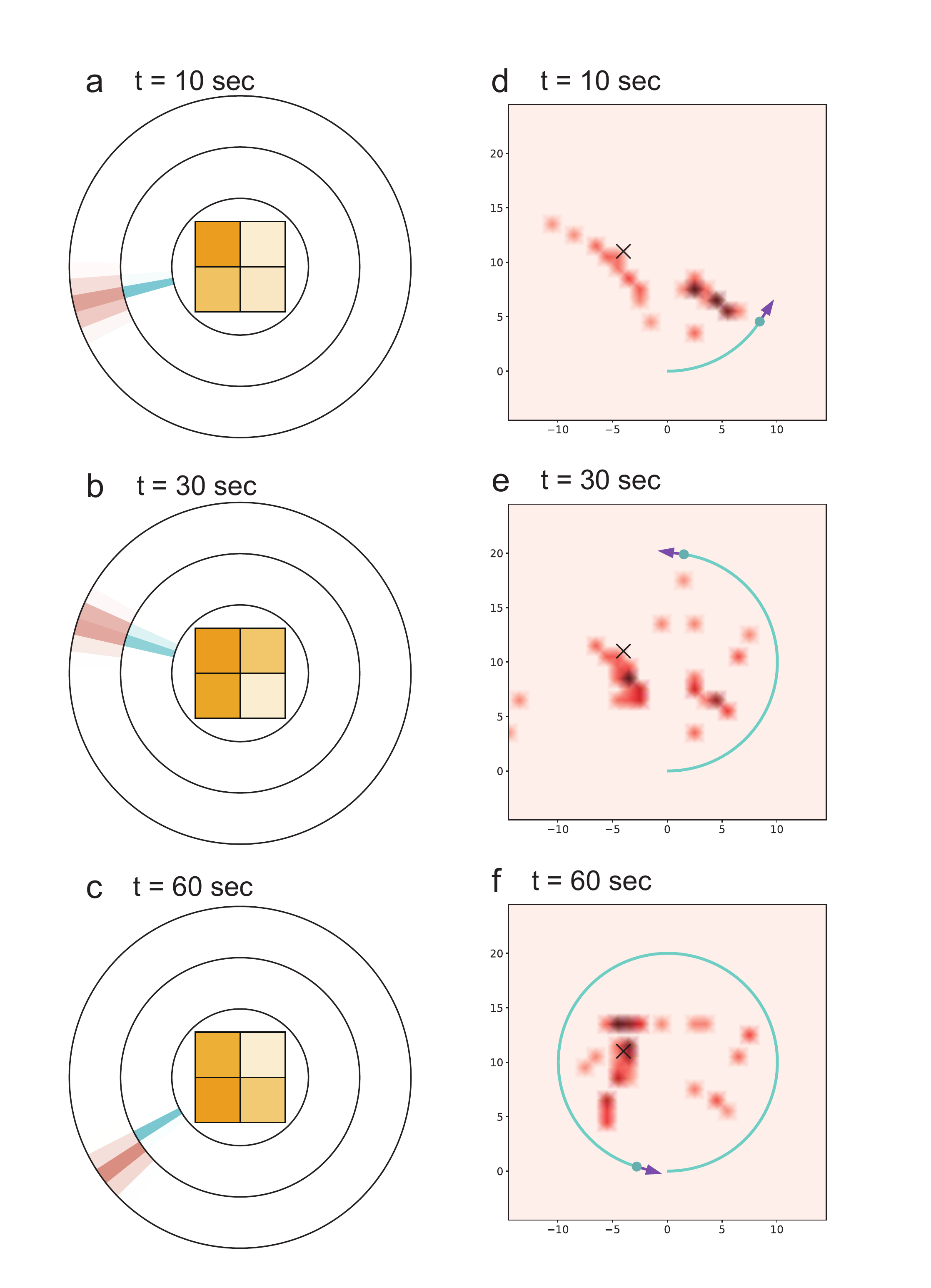}
\caption{\textbf{Radiation mapping with a 2$\times$2 square detector.}   \\
\textbf{a-c.} The detector's input signals and the predicted direction of the radiation source at t=10, 30, 60 sec. The panel with intense colors represent larger value on the detector's signal. On the pie charts the blue and brown sectors represent the ground-truth and predicted direction of radiation source. The top side of the detector in \textbf{a-c.} represents the front side of the moving detector. 
\textbf{d-f.} The process to map the radiation source at t=10, 30, 60 sec. The "$\times$" symbol on the maps show the correct position of the radiation source.  The purple arrows indicate the front side of the moving detector. The area with intense red color indicates the sites with high probability to possess radiation source. The unit of length is the meter.  Check the radiation mapping process in Supplementary Movie 2.
}
\label{SI_map_sq2}
\end{figure}

\begin{figure}[H]
\centering
\includegraphics[height=0.85\textheight]{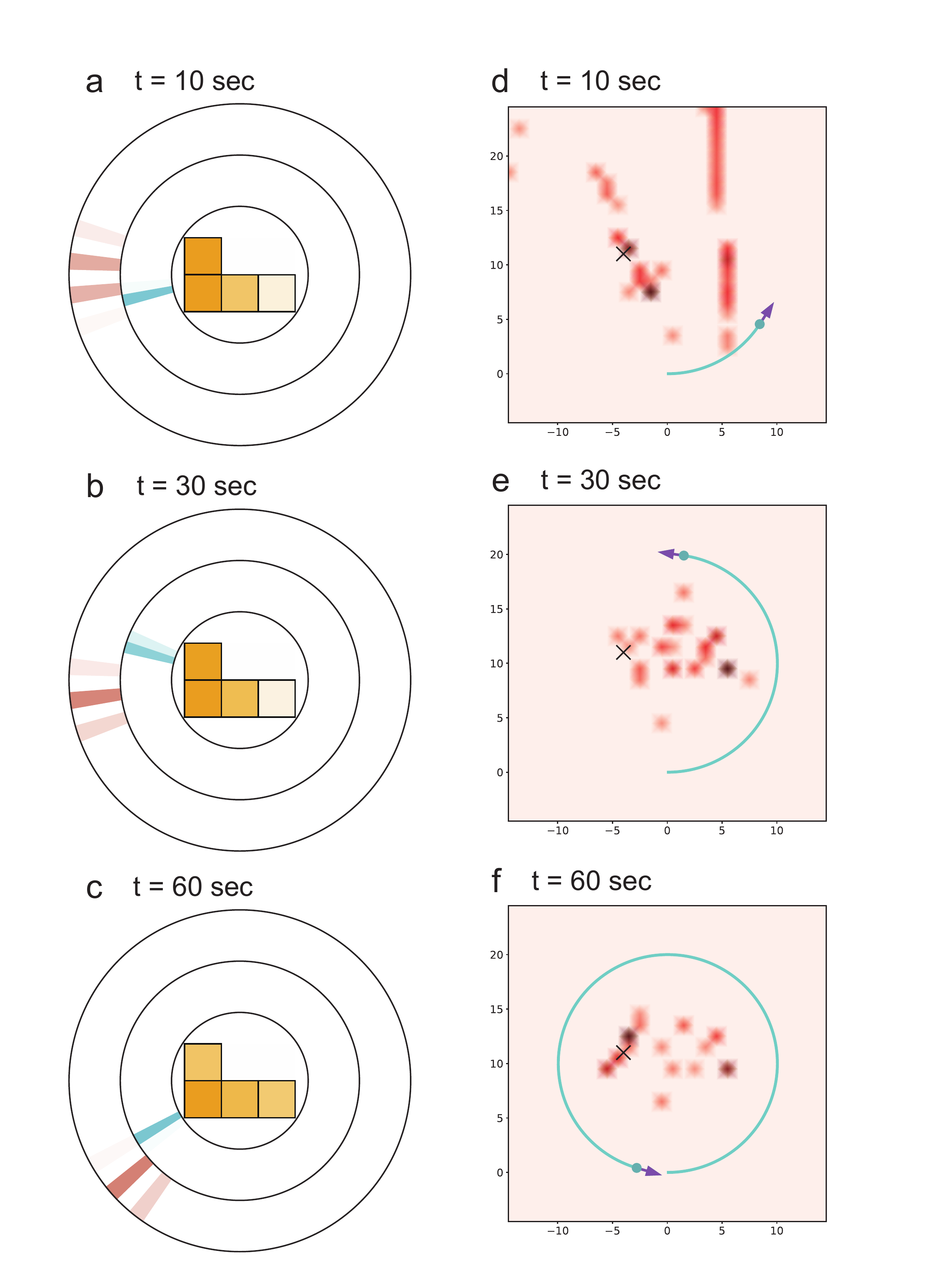}
\caption{\textbf{Radiation mapping with a J-shape Tetris-inspired detector.} \\ 
\textbf{a-c.} The detector's input signals and the predicted direction of the radiation source at t=10, 30, 60 sec. The panel with intense colors represent larger value on the detector's signal. On the pie charts the blue and brown sectors represent the ground-truth and predicted direction of radiation source. The top side of the detector in \textbf{a-c.} represents the front side of the moving detector. 
\textbf{d-f.} The process to map the radiation source at t=10, 30, 60 sec. The "$\times$" symbol on the maps show the correct position of the radiation source.  The purple arrows indicate the front side of the moving detector. The area with intense red color indicates the sites with high probability to possess radiation source. The unit of length is the meter. Check the radiation mapping process in Supplementary Movie 3.
}
\label{SI_map_J}
\end{figure}

\begin{figure}[H]
\centering
\includegraphics[height=0.85\textheight]{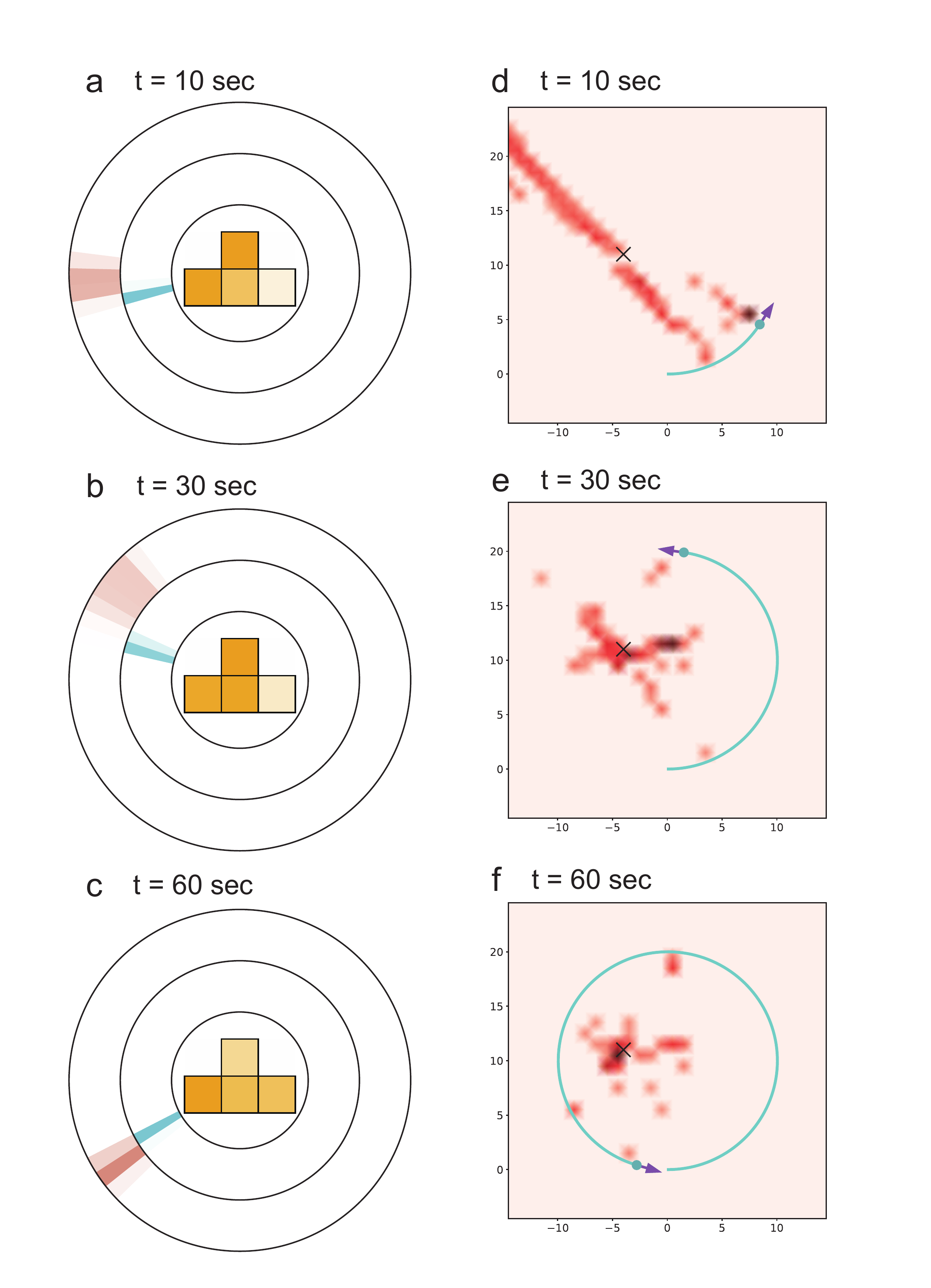}
\caption{\textbf{Radiation mapping with a T-shape Tetris-inspired detector.} \\ 
\textbf{a-c.} The detector's input signals and the predicted direction of the radiation source at t=10, 30, 60 sec. The panel with intense colors represent larger value on the detector's signal. On the pie charts the blue and brown sectors represent the ground-truth and predicted direction of radiation source. The top side of the detector in \textbf{a-c.} represents the front side of the moving detector. 
\textbf{d-f.} The process to map the radiation source at t=10, 30, 60 sec. The "$\times$" symbol on the maps show the correct position of the radiation source.  The purple arrows indicate the front side of the moving detector. The area with intense red color indicates the sites with high probability to possess radiation source. The unit of length is the meter.  Check the radiation mapping process in Supplementary Movie 4.
}
\label{SI_map_T}
\end{figure}

\begin{figure}[H]
\centering
\includegraphics[width=\textwidth]{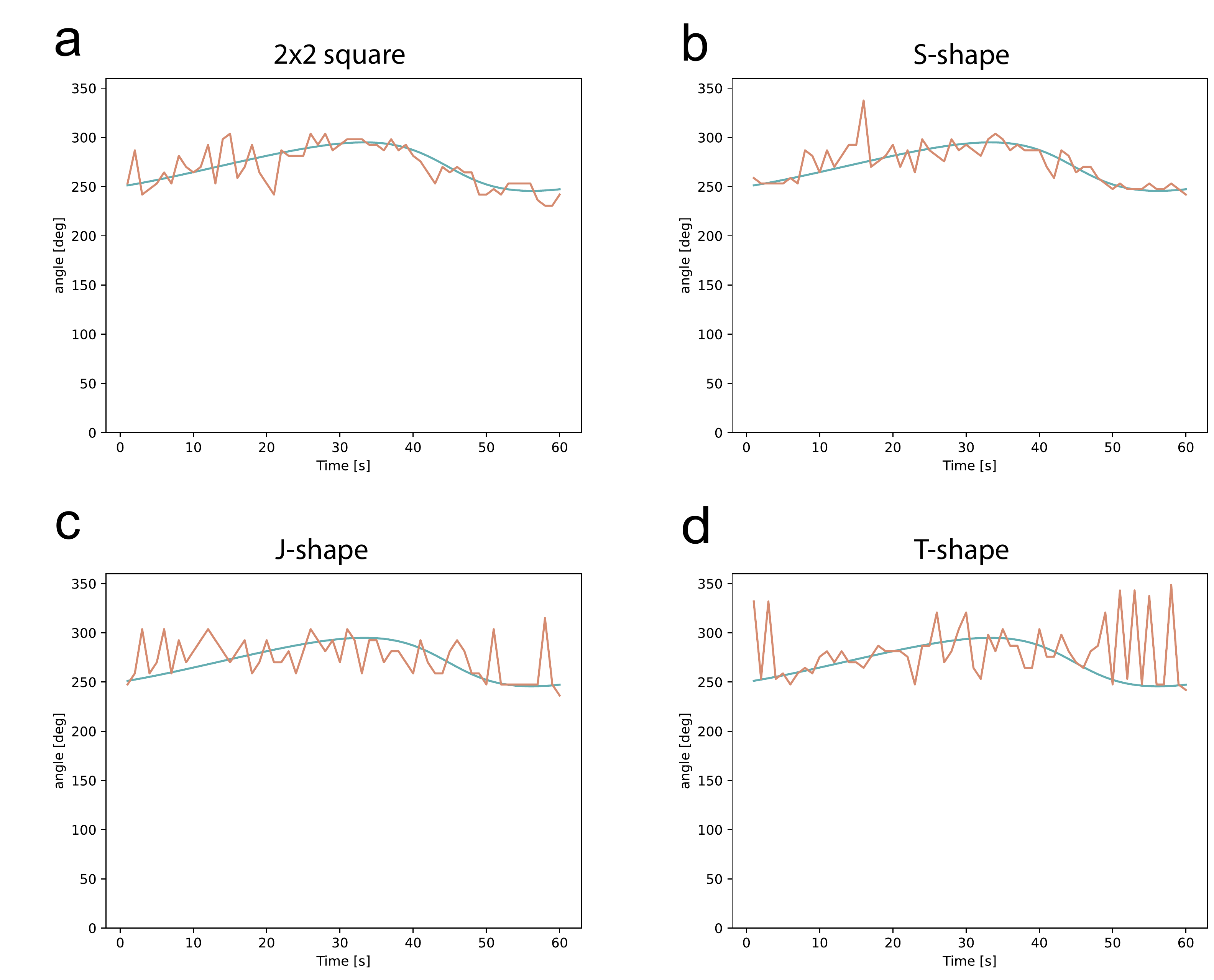}
\caption{\textbf{The predicted and actual directions of the radiation source position.}  \\
The blue and brown lines track the changes of the actual directions and the predicted directions with detectors of \textbf{a.} 2$\times$2 square shape \textbf{b.} S-shape \textbf{c.} J-shape \textbf{d.} T-shape during the radiation mapping process of Figure 3, \ref{SI_map_sq2}-\ref{SI_map_T}. Here the direction is defined as the clockwise angle from the front side of the detector. 
}
\label{SI_angles_1src}
\end{figure}

\begin{figure}[H]
\centering
\includegraphics[height=0.83\textheight]{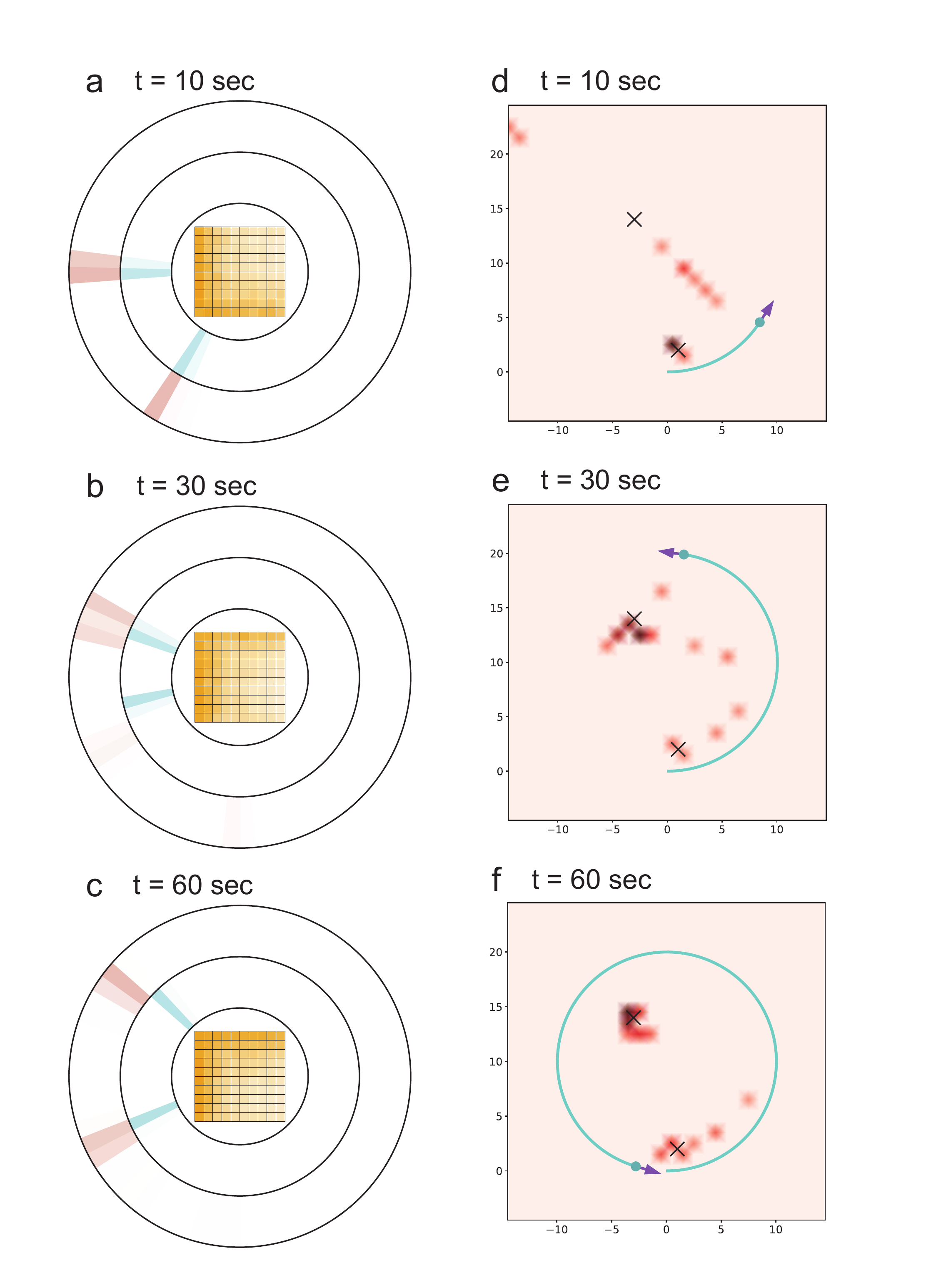}
\caption{\textbf{Mapping 2 radiation sources with a 10$\times$10-size detector.}     \\\
\textbf{a-c.} The detector's input signals and the predicted direction of the radiation source at t=10, 30, 60 sec. The panel with intense colors represent larger value on the detector's signal. On the pie charts the blue and brown sectors represent the ground-truth and predicted direction of radiation source. The top side of the detector in \textbf{a-c.} represents the front side of the moving detector. 
\textbf{d-f.} The process to map the radiation source at t=10, 30, 60 sec. The "$\times$" symbol on the maps show the correct position of the radiation source.  The purple arrows indicate the front side of the moving detector. The area with intense red color indicates the sites with high probability to possess radiation source. The unit of length is the meter. Check the radiation mapping process in Supplementary Movie 5.
}
\label{SI_map_sq10}
\end{figure}

\begin{figure}[H]
\centering
\includegraphics[height=0.83\textheight]{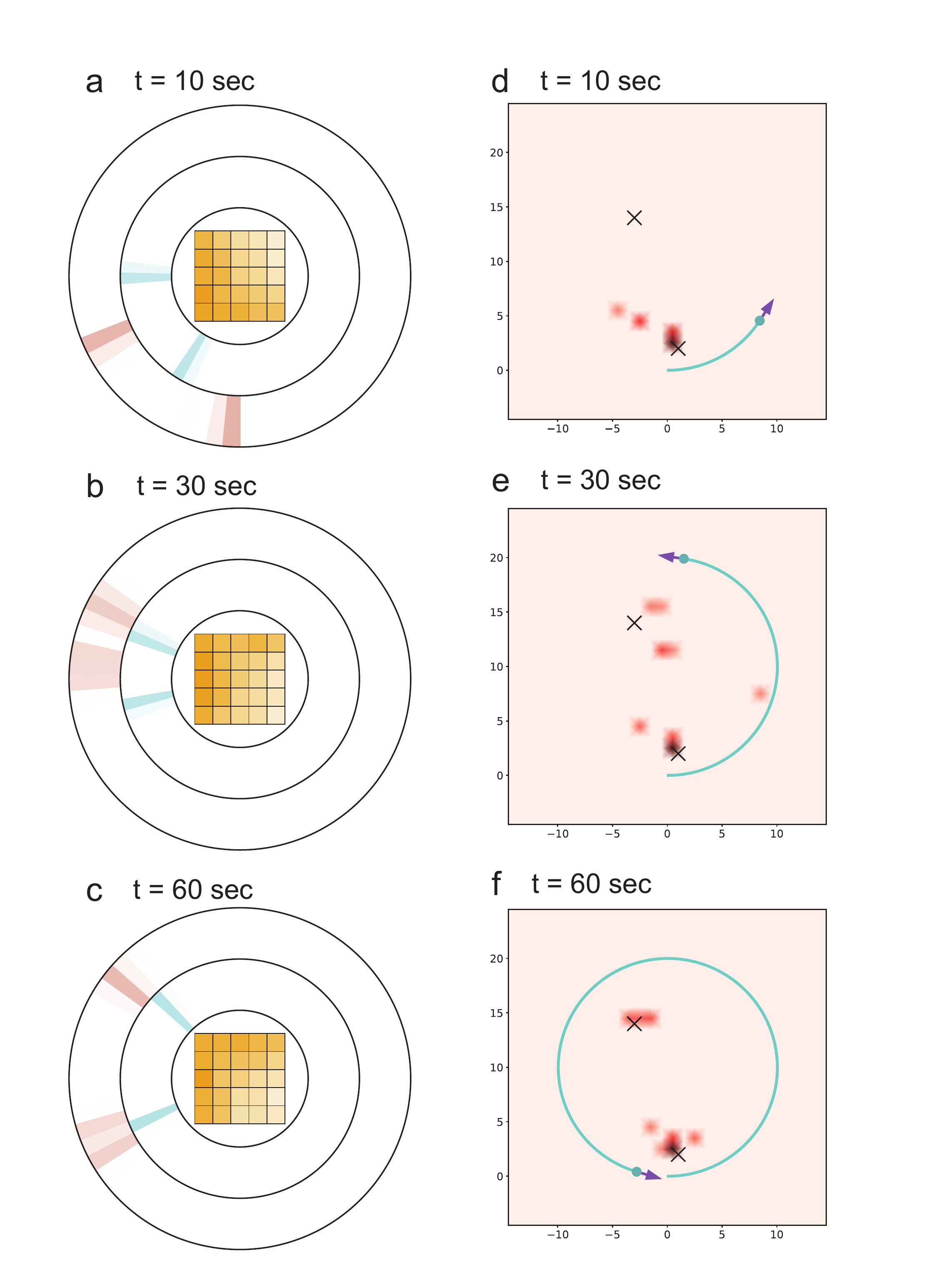}
\caption{\textbf{Mapping 2 radiation sources with a 5$\times$5-size detector.}   \\
\textbf{a-c.} The detector's input signals and the predicted direction of the radiation source at t=10, 30, 60 sec. The panel with intense colors represent larger value on the detector's signal. On the pie charts the blue and brown sectors represent the ground-truth and predicted direction of radiation source. The top side of the detector in \textbf{a-c.} represents the front side of the moving detector. 
\textbf{d-f.} The process to map the radiation source at t=10, 30, 60 sec. The "$\times$" symbol on the maps show the correct position of the radiation source.  The purple arrows indicate the front side of the moving detector. The area with intense red color indicates the sites with high probability to possess radiation source. The unit of length is the meter. Check the radiation mapping process in Supplementary Movie 6.
}
\label{SI_map_sq5}
\end{figure}

A method based on Maximum a Posteriori (MAP) Estimation is applied to generate the radiation distribution map. In the mapping process the intensity contribution of the $i^{th}$ pixel to the $j^{th}$ directional sector of the detector at time $t$ is recorded. Figure \ref{SI_map_explain} explains the roles of these factors in the mapping process. Using this method we have worked on single radiation source detection using detectors with four CZT panels. The result using S-shape detector, which worked the best among the other simple configurations, is shown in our main document. Figure \ref{SI_map_sq2}-\ref{SI_map_T} illustrate the radiation mapping results with 2$\times$2 square, J- and T-shape detectors. To track the directional prediction at each time during the mapping process, we plot the direct predictions of angles in Fig. \ref{SI_angles_1src}. It is obvious that the fluctuation of the directional predictions at each sites reflects the performance of radiation mapping. Furthermore, we have worked on mapping two radiation sources simultaneously. Figure \ref{SI_map_sq10}-\ref{SI_map_sq5} show the result using 10$\times$10 and 5$\times$5 square detector. 

\section{Mapping two radiations sources}

\newpage
\section{Radiation mapping with a rotating detector moving along the circular trajectory.}
\begin{figure}[H]
\centering
\includegraphics[width=0.5\textwidth]{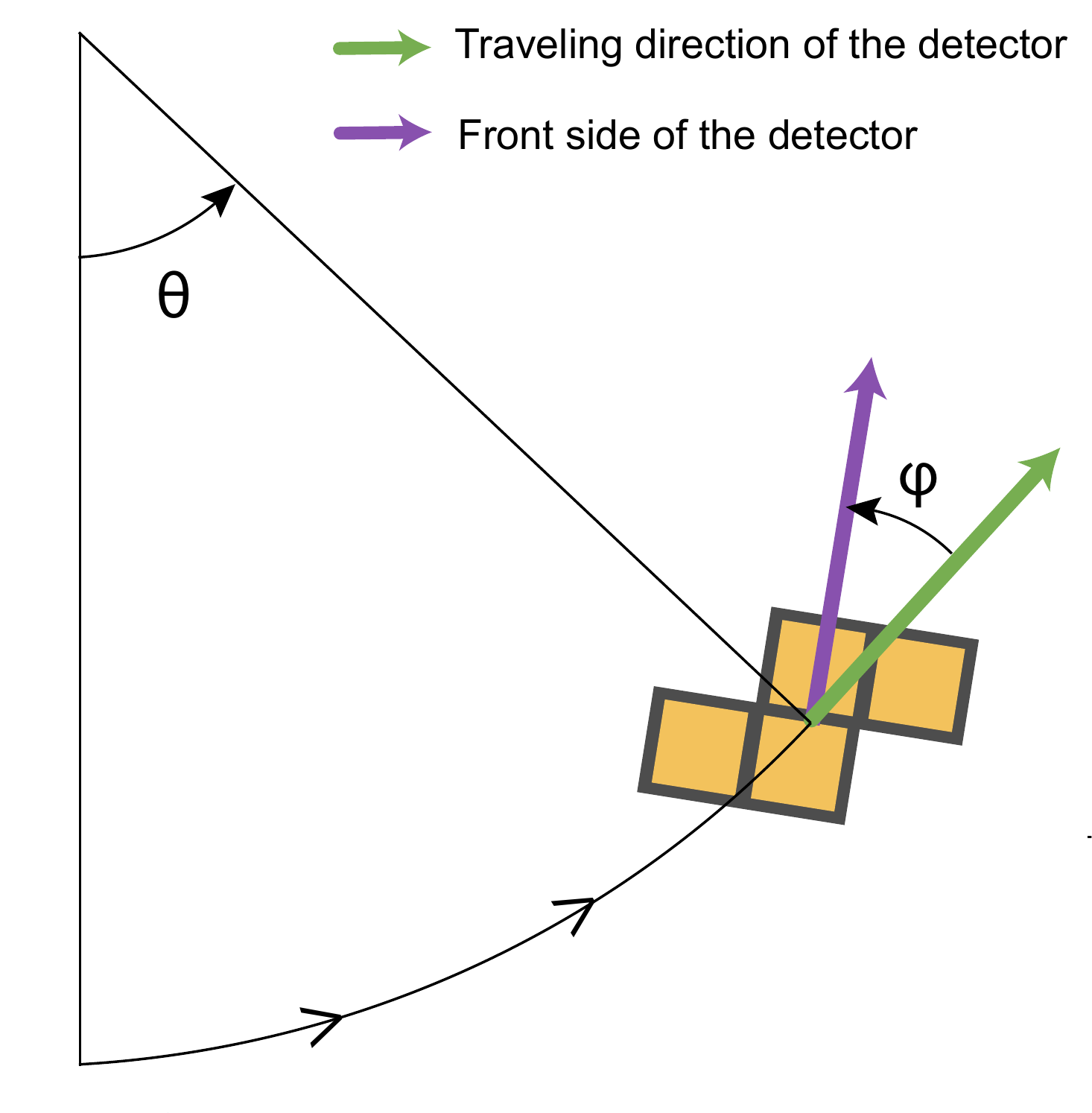}
\caption{\textbf{The method to rotate the directional detector when mapping the radiation source distributions.} 
}
\label{SI_rot_explain}
\end{figure}

\begin{figure}[H]
\centering
\includegraphics[height=0.85\textheight]{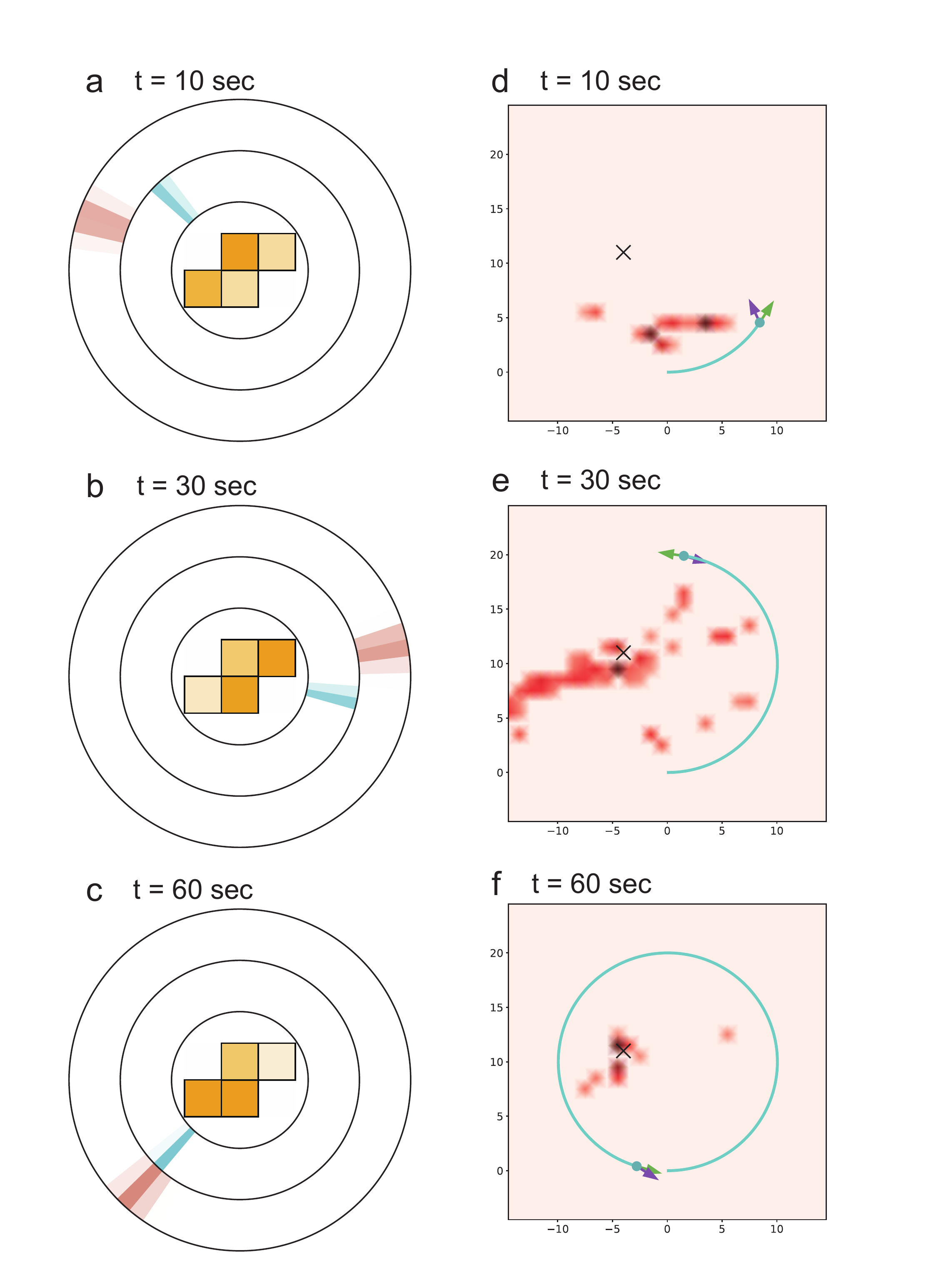}
\caption{\textbf{Radiation mapping with a S-shape Tetris-inspired detector ($\varphi = \theta$).}   \\
\textbf{a-c.} The detector's input signals and the predicted direction of the radiation source at t=10, 30, 60 sec. The panel with intense colors represent larger value on the detector's signal. On the pie charts the blue and brown sectors represent the ground-truth and predicted direction of radiation source. The top side of the detector in \textbf{a-c.} represents the front side of the moving detector. 
\textbf{d-f.} The process to map the radiation source at t=10, 30, 60 sec. The "$\times$" symbol on the maps show the correct position of the radiation source.  The purple and green arrows indicate the front side of the moving detector and the traveling direction respectively. The area with intense red color indicates the sites with high probability to possess radiation source. The unit of length is the meter. Check the radiation mapping process in Supplementary Movie 7.
}
\label{SI_map_S_phi1}
\end{figure}

\begin{figure}[H]
\centering
\includegraphics[height=0.85\textheight]{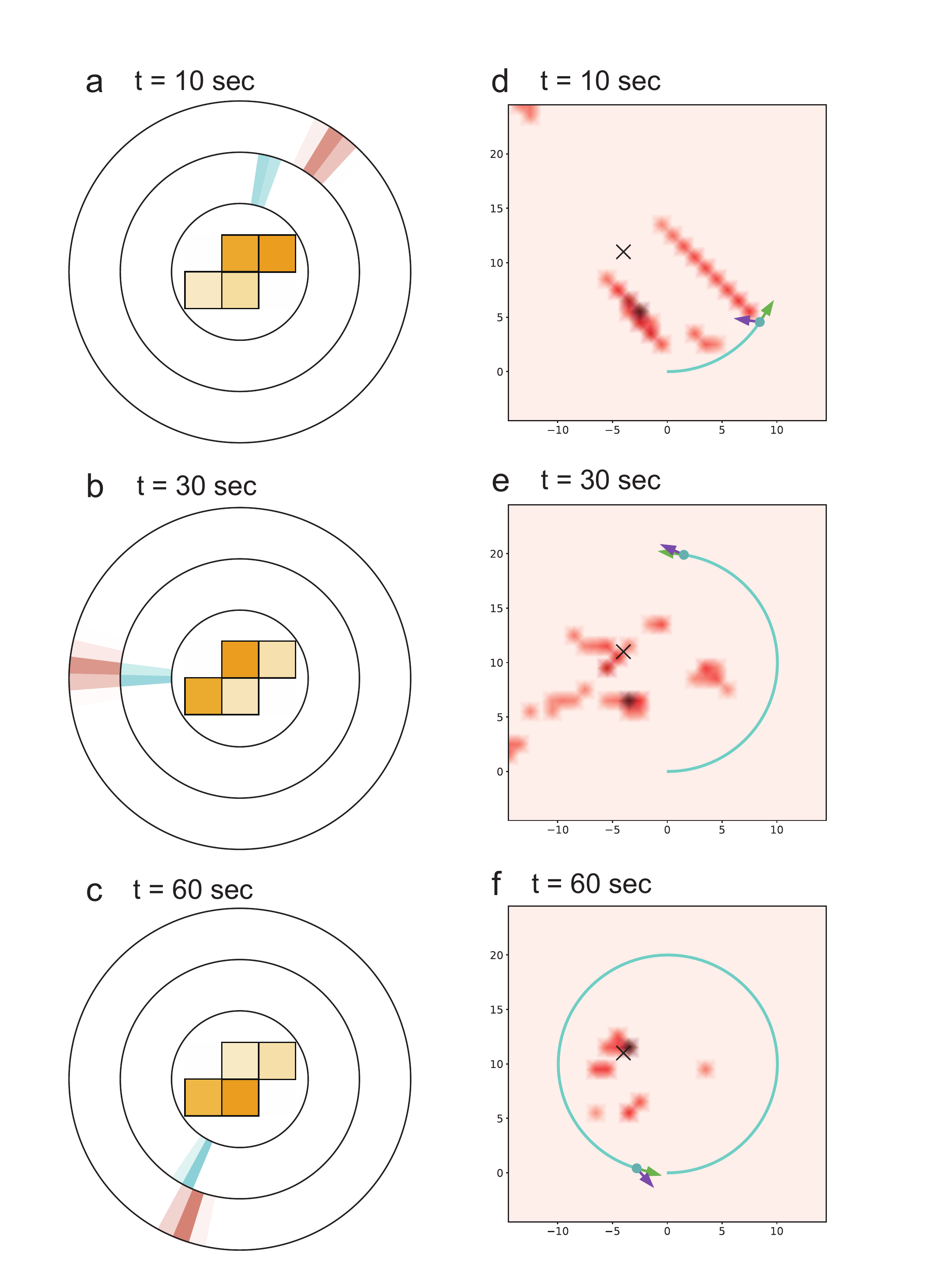}
\caption{\textbf{Radiation mapping with a S-shape Tetris-inspired detector ($\varphi = 2\theta$).}  \\
\textbf{a-c.} The detector's input signals and the predicted direction of the radiation source at t=10, 30, 60 sec. The panel with intense colors represent larger value on the detector's signal. On the pie charts the blue and brown sectors represent the ground-truth and predicted direction of radiation source. The top side of the detector in \textbf{a-c.} represents the front side of the moving detector. 
\textbf{d-f.} The process to map the radiation source at t=10, 30, 60 sec. The "$\times$" symbol on the maps show the correct position of the radiation source.  The purple and green arrows indicate the front side of the moving detector and the traveling direction respectively. The area with intense red color indicates the sites with high probability to possess radiation source. The unit of length is the meter. Check the radiation mapping process in Supplementary Movie 8.
}
\label{SI_map_S_phi2}
\end{figure}

\begin{figure}[H]
\centering
\includegraphics[height=0.85\textheight]{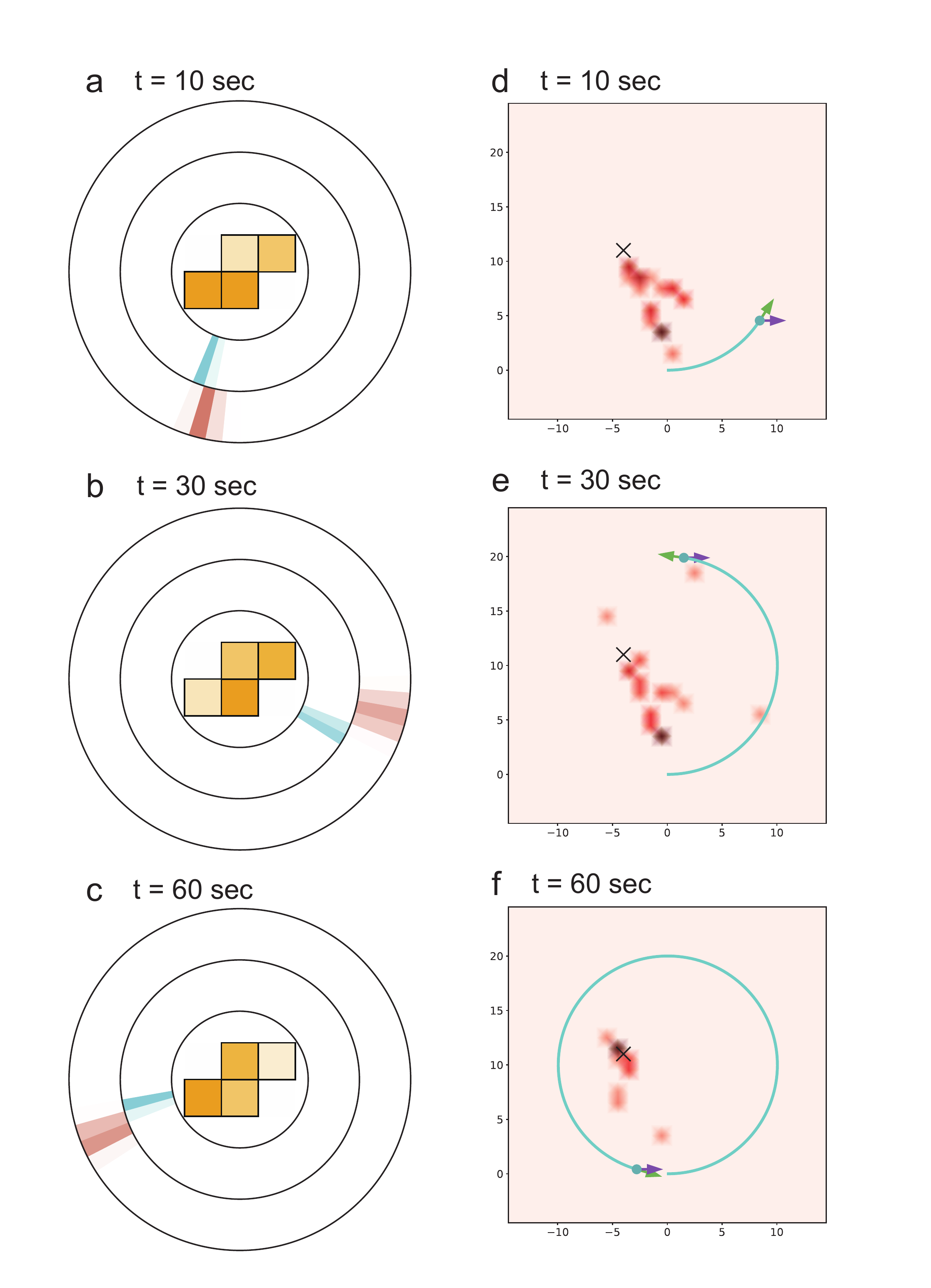}
\caption{\textbf{Radiation mapping with a S-shape Tetris-inspired detector ($\varphi = -\theta$).}   \\
\textbf{a-c.} The detector's input signals and the predicted direction of the radiation source at t=10, 30, 60 sec. The panel with intense colors represent larger value on the detector's signal. On the pie charts the blue and brown sectors represent the ground-truth and predicted direction of radiation source. The top side of the detector in \textbf{a-c.} represents the front side of the moving detector. 
\textbf{d-f.} The process to map the radiation source at t=10, 30, 60 sec. The "$\times$" symbol on the maps show the correct position of the radiation source.  The purple and green arrows indicate the front side of the moving detector and the traveling direction respectively. The area with intense red color indicates the sites with high probability to possess radiation source. The unit of length is the meter. Check the radiation mapping process in Supplementary Movie 9.
}
\label{SI_map_S_phiMinus1}
\end{figure}

\begin{figure}[H]
\centering
\includegraphics[height=0.85\textheight]{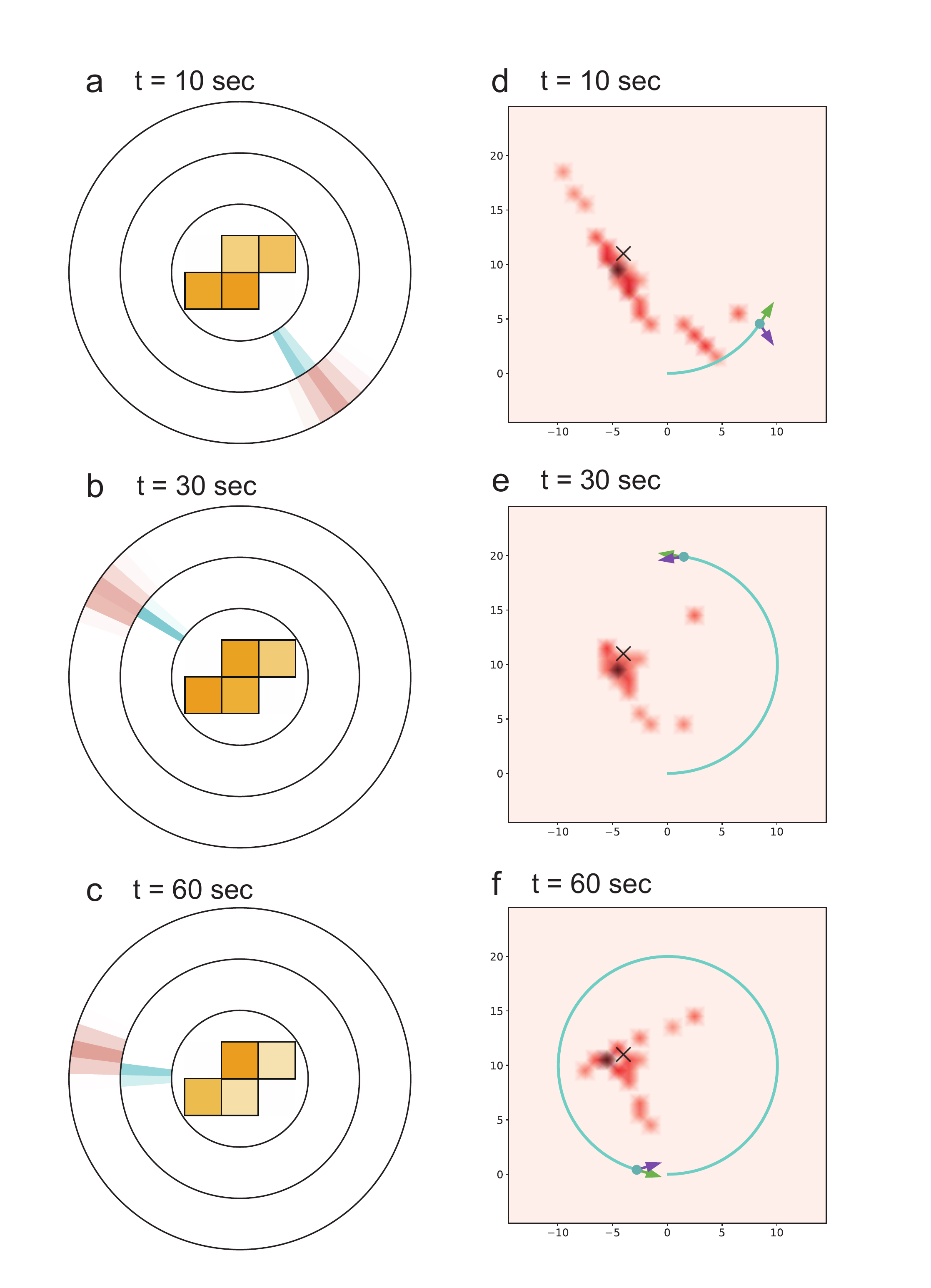}
\caption{\textbf{Radiation mapping with a S-shape Tetris-inspired detector ($\varphi = -2\theta$).}  \\
\textbf{a-c.} The detector's input signals and the predicted direction of the radiation source at t=10, 30, 60 sec. The panel with intense colors represent larger value on the detector's signal. On the pie charts the blue and brown sectors represent the ground-truth and predicted direction of radiation source. The top side of the detector in \textbf{a-c.} represents the front side of the moving detector. 
\textbf{d-f.} The process to map the radiation source at t=10, 30, 60 sec. The "$\times$" symbol on the maps show the correct position of the radiation source.  The purple and green arrows indicate the front side of the moving detector and the traveling direction respectively. The area with intense red color indicates the sites with high probability to possess radiation source. The unit of length is the meter. Check the radiation mapping process in Supplementary Movie 10.
}
\label{SI_map_S_phiMinus2}
\end{figure}

\begin{figure}[H]
\centering
\includegraphics[width=\textwidth]{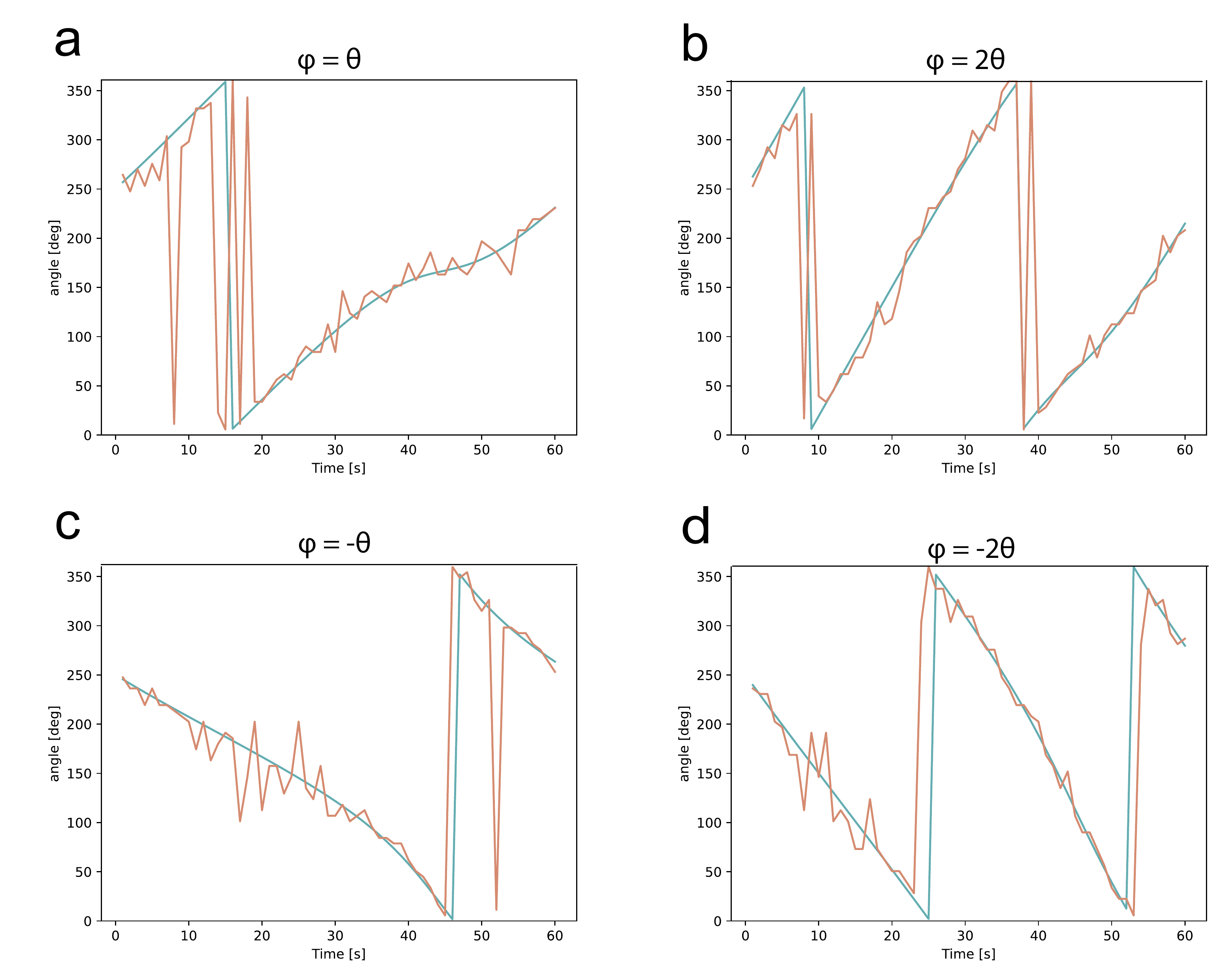}
\caption{\textbf{The predicted and actual directions of the radiation source position.}     \\
The blue and brown lines track the changes of the actual directions and the predicted directions with detectors of \textbf{a.} $\varphi$ = $\theta$, \textbf{b.} $2\theta$, \textbf{c.} $-\theta$, \textbf{d.} $-2\theta$ during the radiation mapping process of Fig. \ref{SI_map_S_phi1}-\ref{SI_map_S_phiMinus2}. Here the direction is defined as the clockwise angle from the front side of the detector. 
}
\label{SI_angles_S_rot}
\end{figure}

In the previous section we have shown that the detectors which are always facing the traveling direction can map radiation sources. However, it is also worthwhile mentioning that the particular detector face that aligns with the detector moving direction does not matter much. The detector facing any direction is already a valid directional detector that is sensitive to radiations coming from all directions. We demonstrate the cases that the moving S-shape detector does not face the traveling direction but rotates at its sites along the circular trajectory. Figure \ref{SI_rot_explain} shows how we set up the simulation. As the detector moves along the circular trajectory by angle $\theta$ anticlockwise, we rotate the detector by $\varphi$ anticlockwise. Figure \ref{SI_map_S_phi1}-\ref{SI_map_S_phiMinus2} show the mapping results with $\varphi$ = $\theta$, $2\theta$, $-\theta$, $-2\theta$. Note that $\varphi=\theta$ means the facing side of the detector looks constant from the static observer. 
We plot the direct predictions of angles in Fig. \ref{SI_angles_S_rot}, as we did in the previous section.  

